\documentclass[twocolumn,prb,showpacs,aps,superscriptaddress]{revtex4-2}

\usepackage{amsmath}
\usepackage{multirow}
\usepackage{graphicx} 
\usepackage{color}
\usepackage[latin1]{inputenc}
\usepackage{enumitem}
\usepackage[french]{babel}

\begin{document}

\date{\today}

\title{Statistics of the cuprate pairon states on a square lattice}

\author{Yves Noat$^*$}

\affiliation{Institut des Nanosciences de Paris, CNRS, UMR 7588 \\
Sorbonne Universit\'{e}, Facult\'{e} des Sciences et Ing\'{e}nierie, 4 place
Jussieu, 75005 Paris, France}

\author{Alain Mauger}

\affiliation{Institut de Min\'{e}ralogie, de Physique des Mat\'{e}riaux et
de Cosmochimie, CNRS, UMR 7590,Sorbonne Universit\'{e}, Facult\'{e} des
Sciences et Ing\'{e}nierie, 4 place Jussieu, 75005 Paris, France}

\author{William Sacks}

\affiliation{Institut de Min\'{e}ralogie, de Physique des Mat\'{e}riaux et
de Cosmochimie, CNRS, UMR 7590,Sorbonne Universit\'{e}, Facult\'{e} des
Sciences et Ing\'{e}nierie, 4 place Jussieu, 75005 Paris, France}

\pacs{74.72.h,74.20.Mn,74.20.Fg}

\pacs{74.72.h,74.20.Mn,74.20.Fg}

\begin{abstract}
In this paper the fundamental parameters of high-$T_c$
superconductivity are shown to be connected to the statistics of
pairons (hole pairs in their antiferromagnetic environment) on a
square lattice.  In particular, we study the density fluctuations
and the distribution of the area surrounding each pairon on the
scale of the antiferromagnetic correlation length $\xi_{AF}$, for
the complete range of hole concentration. We show that the key
parameters of the phase diagram, the $T_c$ dome, and the pseudogap
temperature $T^*$, emerge from the statistical properties of the
pairon disordered state. In this approach, the superconducting and
the pseudogap states appear as inseparable phenomena. The
condensation energy, which fixes the critical temperature, is
directly proportional to the {\it correlation energy} between
pairons and {\it not} to the energy gap, contrary to conventional
superconductors.

When the correlation energy between pairons is suppressed by
fluctuations, either thermally, by disorder, or in the vortex core,
the pseudogap state of disordered pairons is obtained. We attribute
the unique features of cuprate superconductivity to this
order-disorder transition in real space, which clearly differs from
the BCS mechanism. Our predictions are in quantitative agreement
with low-temperature tunneling and photoemission spectroscopy
experiments.
\end{abstract}

\maketitle

\section{Introduction}

Discovered more than 30 years ago by Bednorz and  M\"{u}ller
\cite{ZPhys_Bednorz1986}, high-$T_c$ superconductivity arises in the
cuprate family with critical temperatures going up to 135\,K. The
large variety of compounds have in common the CuO square lattice
which is now accepted to be the active layer where superconductivity
takes place. Despite the wide range of $T_c$
\cite{FrontPhys_Hott2004}, there is a striking universality in the
phase diagram of the various materials. Indeed, as first noted by
Tahir-Kheli et al. \cite{JPhysChem_Tahir2010}, the characteristic
doping points of the $T_c$-dome are essentially material
independent. This fact strongly suggests that a simple mechanism,
directly related to the properties of the CuO planes, governs the
superconducting phase transition and the key aspects
of the physics of cuprates.

In a previous article, using a statistical approach, we showed that the phase diagram of cuprates
is dictated by the topological constraints of the 2D
antiferromagnetic (AF) square lattice on the doped holes
\cite{PhysLettA_Noat2022}. These constraints give rise to two
quantum entities: {\it simplons}, which are individual holes and
{\it pairons}, pairs of holes on neighboring sites, in their local
AF environment \cite{PhysLettA_Noat2022}. Consequently, a simple explanation for the superconducting dome and its
connection to the pseudogap state emerges, wherein both appear as inseparable
phenomena.

In the present work, in addition to the study of the density
fluctuations, we extend the statistical approach to the distribution
of the Voronoi cell areas surrounding each pairon. Let us recall
that the Voronoi cell associated with a given pairon {\it i}
consists of every point in the plane whose distance to the pairon
{\it i} is less than or equal to its distance to any other pairon.
The parameters of the superconducting state, the condensation
energy, and the pairing energy, emerge directly from this analysis.
We propose a phenomenological energy functional, written in terms of
the local density, in which the pairon correlation defines the
condensation energy, and hence the $T_c$. The same energy functional
allows to deduce a novel pairon correlation length, $l_c$, written
in terms of the condensate density, revealing a simple critical
exponent.

\begin{figure}[b]
\includegraphics[width=9.0 cm]{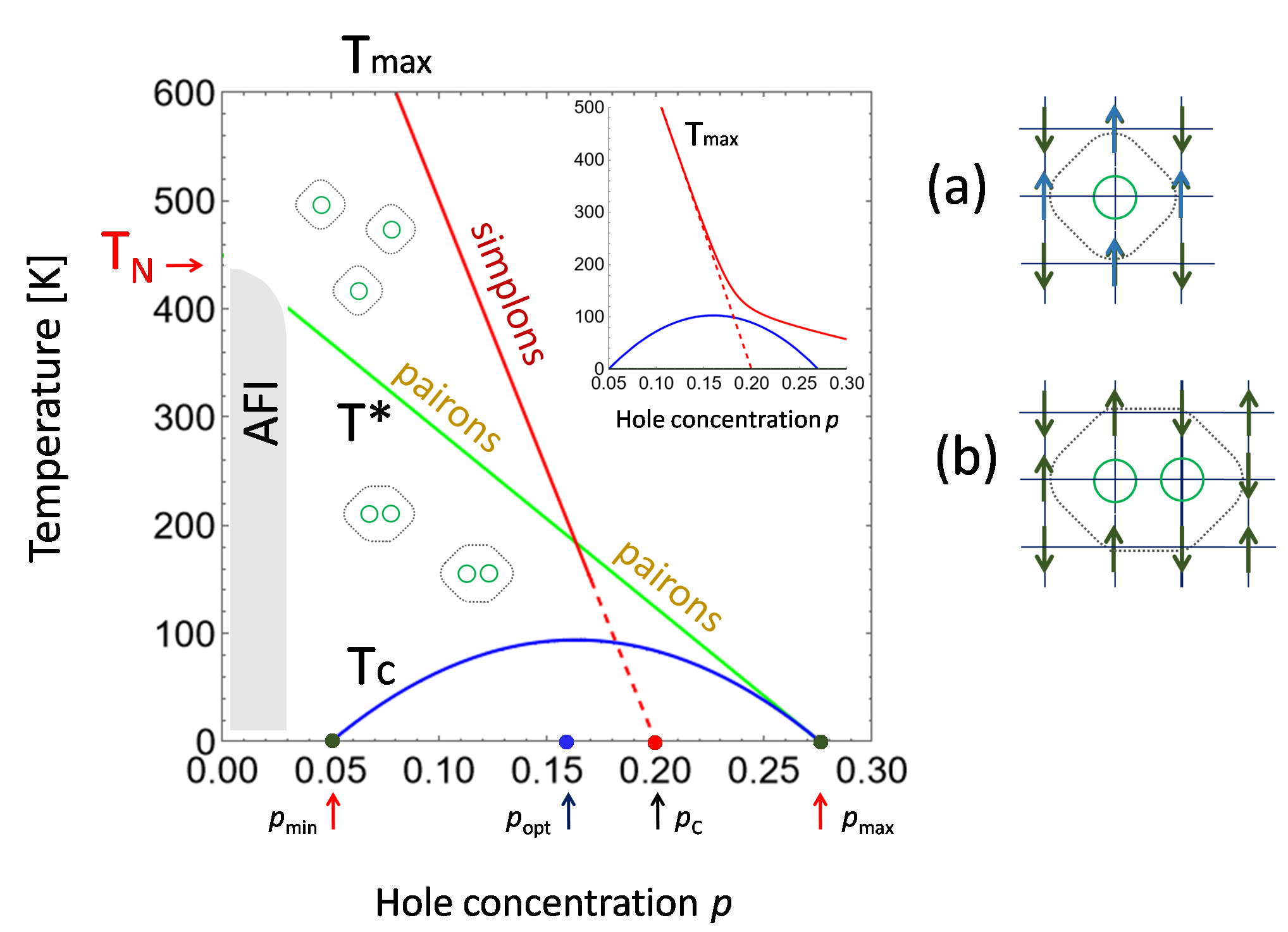}
\caption{(Color online) Schematic phase diagram of cuprates with
important lines: $T_N(p)$, the N\'eel temperature, $T_{max}(p)$, the
temperature of the peak in the magnetic susceptibility, $T^*(p)$ the
pseudogap temperature, $T_c(p)$ the critical temperature. Inset:
Realistic dependence of the magnetic temperature $T_{max}(p)$ (see
text). The latter is linear in the underdoped regime and changes
slope or even saturates in the overdoped regime. Right panel:
Schematic drawing of a simplon (a) and of a pairon (b).}
\label{Fig_Phase diagram}
\end{figure}

The fundamental parameters of the model are deduced quantitatively
from the single energy scale, $J_{eff}$, the effective AF exchange
energy. The estimate of $J_{eff}$ leads to both the $T_c$ dome and
the $T^*$ line as a function of carrier concentration, consistent
with tunneling and photoemission spectroscopy experiments (see
\cite{RepProgPhys_Hufner2008} and references therein). Finally, in
this work, the unconventional cuprate quasiparticle spectrum deduced
from the model is shown to be in quantitative agreement with the
experimental tunneling spectra.

Our work sheds a new light on the nature of the transition in
cuprates, namely a {\it disorder to order} transition of preformed
pairons, which occurs on a characteristic length scale $d_0$, being
the length scale of the pairon-pairon interaction. We assume that
$d_0\sim 6.2 a_0$ ($a_0$ being the lattice constant) is on the order
of the pairon-pairon distance at the onset of the superconducting
dome.


\section{The phase diagram of cuprates: \\ a critical review}

The phase diagram of cuprates is characterized by several
characteristic lines (see Fig.\,\ref{Fig_Phase diagram}). At zero
carrier concentration (hole doping level $p$), the system is an
antiferromagnetic insulator. The N\'eel temperature $T_N(p)$
decreases with $p$ and vanishes at a characteristic value
$p_{AF}\simeq$ 0.02 \cite{PRB_Keimer1992}. For $p\gtrsim p_{min}
\simeq $0.05, the system becomes metallic and a superconductor. As
is well known, the critical temperature $T_c(p)$ follows a parabola
\cite{PhysicaC_Presland1991}: it first increases with doping up to
an optimum value for $p_{opt}$=0.16. It then decreases for larger
doping values and vanishes at the maximum $p_{max} \simeq$ 0.27,
corresponding to the limit of superconductivity. This $T_c$ dome,
firmly established in a wide variety of experiments
\cite{PRL_Torrance_1988,PRB_Takagi1989,
PhysicaC_Presland1991,PRL_Ando2004}, is one of the most fundamental
unanswered properties of cuprates. Surprisingly, its physical origin
has been discussed in relatively few theoretical works, for example
\cite{JPhysChem_Tahir2010,PRB_Feng2012,SUST_Marino2020}.

In the Bardeen-Cooper-Schrieffer (BCS) theory describing
conventional superconductors \cite{PR_BCS1957}, superconductivity
emerges from the metallic state. The formation of the collective
quantum state, a superposition of Cooper pair states
\cite{PR_Cooper1956}, gives rise to a gap $\Delta$ in the single
particle excitation spectrum, which was first evidenced in the
tunneling spectrum by I. Giaever in the 1960s \cite{PR_Giaever1962}.
The spectral gap $\Delta(T)$ is the order parameter; it decreases
with temperature due to fermionic quasiparticle excitations and
finally vanishes at the critical temperature.

Superconductivity in cuprates is unconventional (see Ref.
\cite{PhysicaC_Bosovic2019} and references therein) and deviates
from the BCS theory in many aspects such as\,: the precise shape of
the low-temperature tunneling spectra \cite{Revmod_Fisher2007}, the
unconventional temperature dependence of the specific heat
\cite{PhysicaC_Loram1994,JPhysSocJp_Matsuzaki2004,PRL_Wen2009}, the
anomalous Nernst effect \cite{Nat_Xu2001,PRB_Wang2006}, diamagnetism
\cite{PRB_Li2010} and shot noise above $T_c$ \cite{Nat_Zhou2019},
and finally the non Fermi-liquid normal resistivity
\cite{PRL_Takagi1992,PRL_Ito1993,PRL_Ando2004}.

Contrary to conventional superconductors, in cuprate superconductors
the metallic state is not recovered above $T_c$. Instead, one finds
the well-known pseudogap (PG) state
\cite{Rep_ProgPhys_Timusk1999,LowTempPhys_Kordyuk2015} whose origin
and link to the SC state is still highly debated. The PG state is
characterized by a gap which persists above $T_c$ in the
quasiparticle excitation spectrum up to the higher temperature $T^*$
\cite{PRL_renner1998_T,JphysSocJap_Sekine2016}, while the Josephson
signature of the superconducting coherence still vanishes at the critical
temperature \cite{PRL_Miyakawa1999,JphysConfSer_Sugimoto2021}. Above
$T^*$ one recovers the `normal' metallic state, albeit strongly
correlated.

The onset of a gap in the antinodal direction, as measured by
Angle-Resolved Photoemission Spectroscopy (ARPES)
\cite{Revmod_Damascelli2003,PNAS_Chatterjee2011,JphysSocJp_Yoshida2012,Natcom_Anzai2013,Nat_Hashimoto2014}
defines the pseudogap temperature $T^*(p)$. The latter follows a
straight line as a function of doping $p$
\cite{JPhysSocJap_Nakano1998,RepProgPhys_Hufner2008,PNAS_Chatterjee2011,RepProgPhys_Vishik2018,PRB_Zhong2018}
(see Fig. \ref{Fig_Phase diagram}). As also noted by
\cite{PRB_Cyr-Choiniere2018}, $T^*(p)$  seems to extrapolate to the
N\'eel temperature at zero doping and to vanish at the end of the
dome, i.e. for  $p=p_{max}$.

On the other hand, $T^*$ lines deduced from a variety of experiments
(resistivity
\cite{PhysicaC_Naqib2003,PNAS_Barisic2013,NatCom_Sterpetti2017},
specific heat \cite{PhysicaC_Tallon2001}, magnetic susceptibility
\cite{ScSciTech_Naqib2008}) have been found to contradict the above
scenario. These lines are suggested to cross the dome and vanish at
a critical point $p\sim 0.2$ \cite{PhysSolStat_Tallon1999}. This
conundrum, still debated, was answered in our previous work in
connection to simplons \cite{Solstatcom_Noat2022} and will be
further substantiated below.


The magnetic susceptibility of cuprates has been extensively studied
and exhibits a broad peak as a function of temperature
\cite{PRL_Johnston1989,PRB_Torrance1989,PhysicaC_Yoshizaki1990,PhysicaC_Oda1991,PRB_Nakano1994}.
The position of the peak, defined as $T_{max}$, reveals the
characteristic temperature of magnetic correlations \cite{JphysChemSol_Lines1970}. Well above
$T_{max}$, the magnetic susceptibility follows a Curie-Weiss law of
quasi-independent spins, while for $T<T_{max}$, the magnetic
susceptibility decreases upon cooling due to growing magnetic
correlations. So $T_{max}(p)$ represents the crossover between these two regimes.

In the hole doping phase diagram, $T_{max}(p)$ decreases linearly
as a function of doping and then changes slope or even saturates in
the overdoped regime \cite{PRB_Nakano1994,Solstatcom_Noat2022}, as
illustrated in the inset of Fig. \ref{Fig_Phase diagram}. One
remarks that $T_{max}(p)$, for a wide range of $p$, is notably
larger than $T^*(p)$, indicating that the two phenomena are
distinct. Indeed, at the SC onset ($p \simeq .05$), for
Bi$_2$Sr$_2$CaCu$_2$O$_{8+\delta}$ (BSCCO$_{2212}$), the two
temperatures are roughly in a 2 to 1 ratio.


\subsection{Simplons and pairons}

In our approach, the phase diagram of cuprates can be understood
quantitatively by considering two quantum entities, simplons and
pairons, which directly result from hole doping of the CuO square
lattice \cite{Solstatcom_Noat2022}.

The simplon is a hole surrounded by four `frozen' spins (see Fig.
\ref{Fig_Phase diagram}, upper right panel), which we introduced in
a previous work in order to understand the behavior of the
$T_{max}(p)$ line deduced from the magnetic susceptibility
\cite{Solstatcom_Noat2022}. A hole in the CuO lattice evidently
suppresses one spin, but when magnetic correlations are strong
enough a `simplon' is formed. The latter has the property of one
central spin plus 4 frozen spins on nearest neighbor sites, thus the
suppression of fluctuations of 5 spins per hole. The necessary
condition for simplon formation is that the correlation length
satisfies $\xi_{AF}(T)\gtrsim 2\, a_0$, where $a_0$ is the lattice
parameter, which is true for temperatures such that $T\lesssim
T_{max}$.

A simplon is related to a skyrmion but with a finite correlation
length, since a skyrmion also freezes the 4 spins in the CuO plane
around the hole. At very low doping ($p<p_{AF}$), before metallicity
is established, a simple picture is that one hole creates a simplon
below $T_{max}$ which evolves into a skyrmion at lower temperature
when magnetic correlations extend to a much larger scale such that
$\xi_{AF}(T)\gg 2 a_0$. Moreover, the formation of skyrmions is a
realistic scenario to explain the disappearance of long range AF
order at low doping \cite{JPhysSocJp_Suda2016}.

In the metallic region, i.e. for $p>p_{min}$, we proposed that holes
(simplons) combine to form hole pairs or pairons
\cite{EPL_Sacks2017} below the characteristic temperature $T^*$.
This temperature defines the pseudogap phase by the onset of the
spectral gap $\Delta_p$ in the antinodal direction, which can be
precisely measured by ARPES
\cite{Nat_Hashimoto2014,Natcom_Anzai2013, RepProgPhys_Vishik2018}.
In our approach, the binding energy of a pairon is determined by the
surface area of the local AF island surrounding the two holes on the
characteristic length scale $\xi_{AF}$ \cite{EPL_Sacks2017}, which
was found to vary as the inverse distance between holes $\sim
1/\sqrt{p}$ \cite{PRB_Birgeneau1988}. Consequently, it is natural
that $T^*$ decreases with increasing hole doping and vanishes at
$p_{max}$, a critical value which is determined by the minimum area
of a pairon $A_c$ \cite{EPL_Sacks2017}:
\begin{equation}
\Delta_p(p) \simeq 2.2\, k_B T^*\simeq J_{eff}
\left(1-\frac{p-p_{min}}{p_{max}-p_{min}}\right) \label{Equa_Tstar}
\end{equation}
with $p_{max} \simeq 2\,a_0^2/A_c\simeq 0.27$.

As we will see in a later section, this linear
dependence of the antinodal gap is remarkably well satisfied. Here
$J_{eff}$ is conveniently defined as the effective exchange energy
at the value $p = p_{min}$, about $\sim 74$ meV for BSCCO$_{2212}$.
Moreover, the fact that $\Delta_p(p) \propto J_{eff}$ indicates that
the pairing mechanism is linked to the local magnetism in a static,
and not retarded, interaction of coherent spins on the scale of
$\xi_{AF}$, the AF coherence length.

\begin{figure}
\includegraphics[width=8.4 cm]{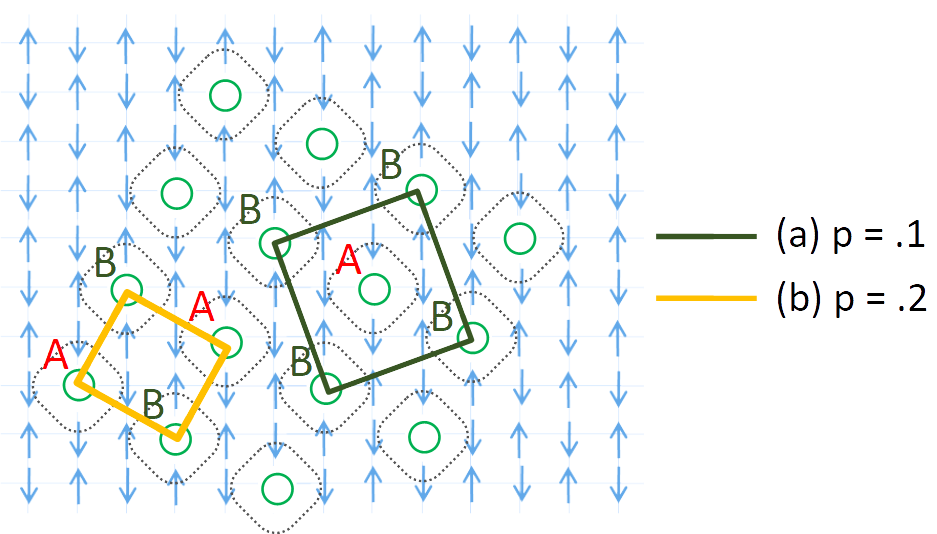}
\caption{(Color online) Compact network of simplons. Yellow square:
elementary unit cell of the compact simplon lattice, at the doping
value $p=$0.2 corresponding to the critical value $p_c$ in Fig.
\ref{Fig_Phase diagram}. Green square: elementary unit cell of the
compact lattice of equivalent simplons, corresponding to the doping
value $p=$0.1.} \label{Fig_Simplons}
\end{figure}

\subsection{Magnetic temperature scale}

In the early days of high-$T_c$, the magnetic susceptibility,
$\chi(T)$, was studied for a wide range of carrier concentrations in
several materials. Moreover, theoretical calculations were
successful to describe many of the magnetic properties
\cite{RevModPhys_Mabousakis1991}, in contrast to the SC issues.

As mentioned previously, the position of the peak in the
susceptibility, $T_{max}(p)$, decreases with hole doping in a
reproducible fashion (see Fig. \ref{Fig_Phase diagram})
\cite{PRB_Nakano1994,Solstatcom_Noat2022}. For decades, many authors
confused $T_{max}$ with $T^*$ (see Ref. \cite{Solstatcom_Noat2022}
for further details). This led to contradicting definitions of the
pseudogap region in the phase diagram. Recently, we reexamined the
susceptibility data for different materials and found that the two
temperature scales $T^*$ and $T_{max}$ are clearly distinct,
allowing to resolve this paradox \cite{Solstatcom_Noat2022}. In
agreement with early work by Nakano et al. \cite{PRB_Nakano1994},
$T_{max}(p)$ first decreases linearly with increasing $p$ over a
wide range, but then softens in slope in the overdoped
regime\,\cite{Solstatcom_Noat2022} (inset of Fig.\ref{Fig_Phase
diagram}).

The formation of simplons provides a very simple explanation for the
linear behavior of $T_{max}(p)$. In this picture, a hole freezes
completely the four nearest neighboring spins. Consequently, a total
of 5 spins per hole no longer contributes to the magnetic
susceptibility. It follows that\,:
\begin{equation}
k_B\,T_{max}(p) \simeq J (1-5\,p)
\end{equation}
where $J$ is the intrinsic exchange interaction of the CuO plane
(for BSCCO$_{2212}$, $J \sim 1.3\, J_{eff}$).

Consequently, the linear part of $T_{max}(p)$ extrapolates down to
zero at the characteristic value $p_c = $0.2, which is close to a
possible quantum critical point \cite{AnRevCondMat_Proust2019}. As
discussed in our previous paper \cite{PhysLettA_Noat2022}, the
linear extrapolation to $p_c$ is a consequence of simplons in the
high density limit. Indeed, the concentration $p_c = 1/5$
corresponds to the perfect compact simplon lattice, as illustrated
in Fig. \ref{Fig_Simplons}. At the
extrapolated density of this compact simplon lattice, all spins of
the system are frozen.

\begin{figure*}[t]
\centering
\includegraphics[width=18 cm]{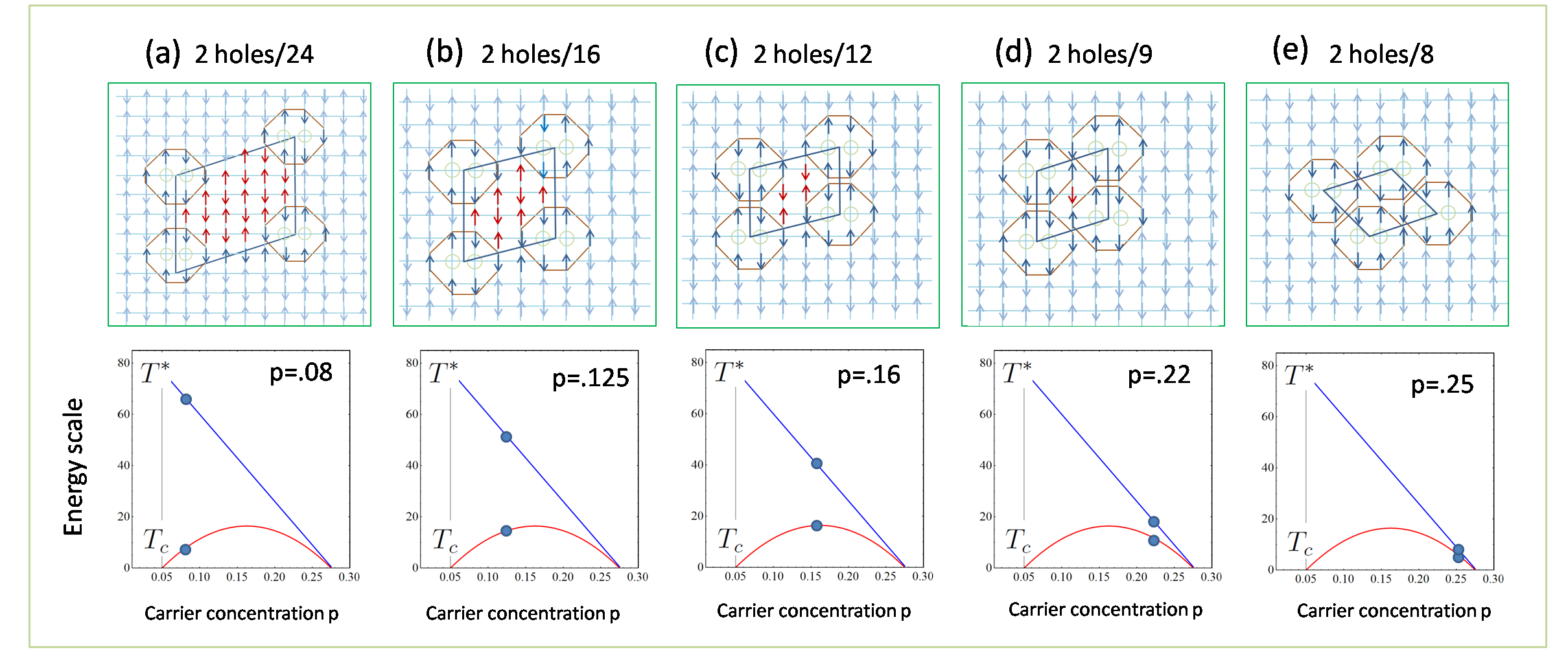}
\caption{(Color online) Pairon network for different doping values:
a) underdoped regime $p=$0.08; b) $p=$0.125; c) optimally doped
$p=$0.16; d) overdoped regime $p=$0.22; e) Maximum doping (compact
lattice) $p=$0.25. In each case, the AF spin correlation length, the
average distance between pairons, can be estimated using $\xi_{AF}
\simeq \sqrt{2/p}$ \,: (a) $\xi_{AF}\sim 5 a_0$, (b) $\xi_{AF}\sim 4
a_0$, (c) $\xi_{AF}\sim 3.5 a_0$, (d) $\xi_{AF}\sim 3 a_0$, (e)
$\xi_{AF}\sim 2.8 a_0$. The pairon-pairon interaction threshold, at
the onset of superconductivity, occurs at $p\simeq .05$ where
$\xi_{AF}\sim 6.3 a_0$.}. \label{Fig_Networks}
\end{figure*}

It is worth noting that the values of $p_c$ vary slightly from one
compound to the other \cite{Solstatcom_Noat2022}. A possible
explanation is the influence of interlayer coupling which reduces
spin fluctuations and enhances the N\'eel temperature $T_N$ at zero
doping \cite{PRB_Singh1992}. The interlayer coupling is stronger for
Bi$_2$Sr$_2$CaCu$_2$O$_{8+\delta}$ (BSCCO$_{2212}$) than for
La$_{2-x}$Sr$_x$CuO$_4$ (LSCO), the latter being more 2-dimensional
with a lower $T_N$. As a result, the four neighboring spins could be
only partially frozen in LSCO, thus giving a smaller slope and a
larger $p_c\approx 0.23$ than for
Y$_{1-x}$Ca$_x$Ba$_2$Cu$_3$O$_{7+\delta}$ (YBCO) or BSCCO$_{2212}$
for which $p_c\approx 0.20$.

As seen in the phase diagram, the two characteristic temperatures
approach each other in the overdoped regime. This might explain why
$T_{max}(p)$ line deviates from linearity. Indeed, below $T^*$,
holes bind together to form pairons, which affects the magnetic
correlations in a different way than simplons. As seen in the phase
diagram the smooth evolution of $T_{max}(p)$ extends to the limit of
the dome, suggesting that magnetic correlations survive up to
$p_{max}$ \cite{PRB_LeTacon2013,PRB_Peng2018} or even beyond, as
seen in some experiments \cite{Nature_Dean2013}.

\section{Statistics of pairon states}

\subsection{Fluctuation of the density}

We now study the statistical properties of pairons randomly
distributed on a 2D square lattice. As described below, this state
can be characterized either by studying the local density or by the
distribution of the Voronoi cell areas. We recall that the Voronoi
cell is the generalization of the Wigner-Seitz cell for a non
periodic distribution of points. We make the linear transformation
for the doping in such a way that $p^\prime=0$ corresponds to the
onset of superconductivity and that $p^\prime=1$ corresponds to the
compact pairon lattice \cite{PhysLettA_Noat2022}, which is obtained
for $p=$0.25 (see Fig. \ref{Fig_Networks}). The latter value is very
close to the end of the SC dome (i.e. $p=p_{max}$) :
\begin{equation}
p^\prime=\frac{p-p_{min}}{p_{max}-p_{min}}
\end{equation}

\begin{figure}
\includegraphics[width=9 cm]{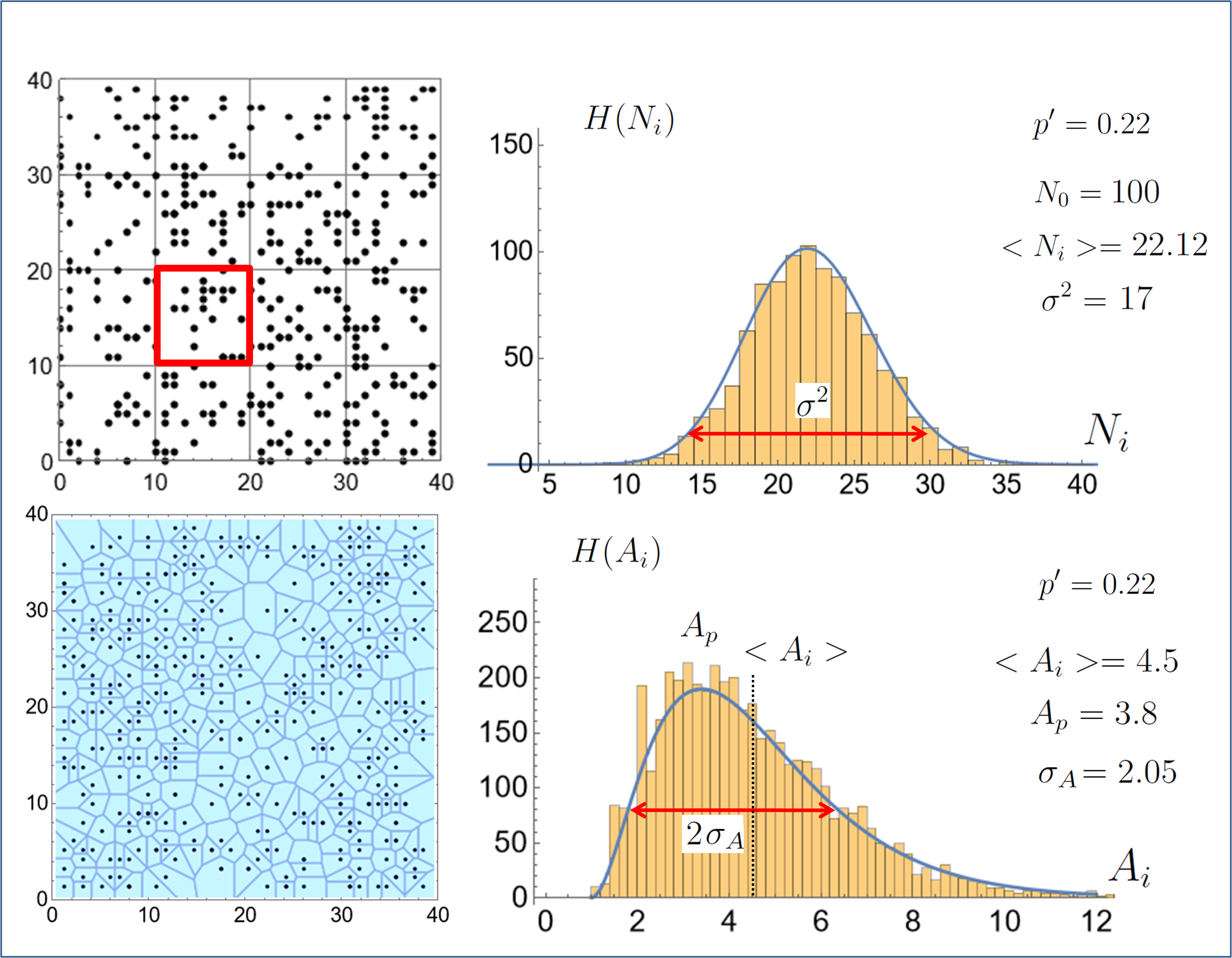}
\caption{(Color online) Upper panel: Statistics of the number of
pairons randomly distributed on a square lattice with 40$\times $40
sites. We count the number of pairons $N_i$ found inside the square
of side $d_0=10$ for each configuration and take the average over
the configurations. Solid line is the exact binomial law curve
corresponding to $p\prime=0.22$. Lower panel: Statistics of the
Voronoi cells calculated for a 40$\times$40 lattice. Solid line is
the distribution curve given by Eq. \ref{Equa_Distrib_Voro}.}
\label{Fig_Statistics}
\end{figure}

The numerical simulation proceeds as follows. We
consider a discrete square lattice of pairon sites of size $N_T
\times N_T$, where in practice $N_T$ = 40
(Fig. \ref{Fig_Statistics}). Then, we distribute randomly a fixed number of pairons
on the $N_T \times N_T$ sites, with the only constraint that two
pairons cannot occupy the same site. By construct, a given site has
the probability $p'$ of being occupied and $1-p'$ of being vacant.

The statistics of the pairons is then calculated in
a smaller cell within the larger square\,: a cell of area $d_0
\times d_0$, containing $N_0$ sites, to avoid edge effects (see
Fig. \ref{Fig_Statistics} upper panel). For convenience, we chose
$d_0=10$, and define $N_i$ the number of pairons in such a square in
the $i$th configuration. We then consider successive random
configurations to study their statistical distribution, keeping
constant the total number of pairons in the larger square of size
$N_T \times N_T$. After a sufficiently large number of trials, we
check that the average number of the $N_i$ is independent of $i$ to
the desired accuracy.

By the previous definitions, $N_i/N_0$ is the local pairon density
of the $i$th configuration and is obviously proportional to the
filling factor $p^\prime$:
\begin{equation}
\left\langle N_i\right\rangle=N_0\,p^\prime \label{Equa_Moy}
\end{equation}
The variance $\sigma^2=\left\langle N_i^2\right\rangle-\left\langle
N_i\right\rangle^2$ follows the binomial law:
\begin{equation}
\sigma^2=N_0\,p^\prime (1-p^\prime) \label{Equa_Variance}
\end{equation}
The result is illustrated in Fig. \ref{Fig_Statistics}, upper panel. The $ N_i$
distribution is calculated for a 40$\times$40 lattice for $d_0=10$
and doping $p^\prime=$0.22 and compared to the theoretical binomial
distribution curve for the corresponding density (blue line).

These results confirm that $\sigma^2$ characterizes the strength of
the disorder and follows the expected binomial law. We proposed in a
previous article \cite{PhysLettA_Noat2022} that the variance
$\sigma^2$ is proportional to the work needed to disorder the system
at zero temperature, starting from the (ordered) SC state, to reach
the disordered Cooper Glass State (CPG). The latter is an incoherent
non-SC state, wherein the system is totally disordered, the variance
is a maximum, reaching the binomial value (Eq. \ref{Equa_Variance}).
Therefore, by definition, no more correlations between pairons
exist. The work $W\propto \sigma^2$ needed to go from the SC state
to the CPG state represents the coherence energy $\beta_c$ of the
system which, in our view, is directly proportional to the critical
temperature \cite{SciTech_Sacks2015,EPL_Sacks2017}.

The SC to CPG transformation explains the origin of the $T_c$-dome.
Since $$T_c(p)\propto \sigma^2\propto p^\prime (1-p^\prime)$$ it is
exactly an inverted parabola, as dictated by the binomial law. The
critical temperature is maximum at the optimum doping $p^\prime$=0.5
(equivalent to $p=0.16$), for which the disorder strength is maximum.
For $p^\prime>0.5$, on the overdoped side, the number of vacant sites
decreases with $p^\prime$ so the disorder strength becomes progressively
lower and finally, for $p^\prime =1$, all pairon sites are occupied
(see Fig.\ref{Fig_Networks}), there is no energy gain at all, and
$\beta_c = 0$. At this critical density, the mean gap $\Delta_p$
also vanishes (see Eq.\ref{Equa_Tstar}) so the two important energy scales,
$\beta_c$ and $\Delta_p$, are both zero at the end of the SC dome.

\begin{figure}[h!]
\includegraphics[width=7 cm]{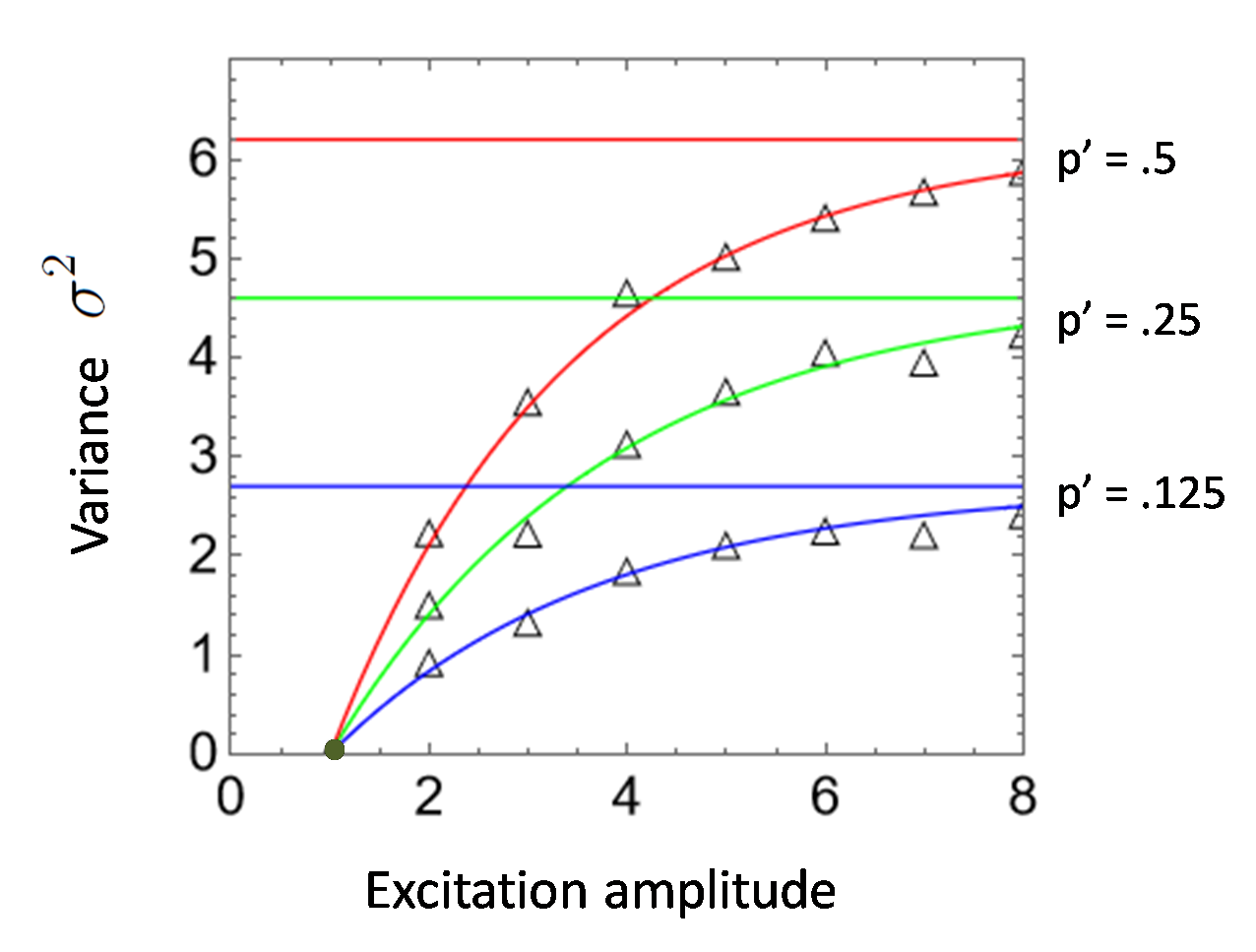}
\caption{(Color online). Classical model for the order-disorder
transition starting from a pairon Wigner lattice subject to a
stochastic noise of increasing amplitude. The variance $\sigma^2$ of
the pairon distribution is plotted for three pairon densities $p'$
as a function of noise amplitude (in units of the pairon lattice).
The variance increases monotonically and finally reaches the
binomial law value, as expected. The curved lines serve as a guide
to the eye.} \label{Fig_variance}
\end{figure}

To illustrate the adiabatic transformation from ordered to
disordered states, we consider a classical model wherein the initial
ordered state is a perfect lattice of pairons, analogous to the
Wigner crystal of classical interacting charged  beads
\cite{EPL_StJean2001}. The lattice is then subject to a stochastic
noise of increasing amplitude. The results for $\sigma^2$ ($\propto
W$) are plotted in Fig. \ref{Fig_variance} for three different
pairon densities. One can see clearly that the variance increases
monotonically with noise amplitude, until reaching the completely
disordered state verifying the binomial law $\sigma^2 = N_0\,
p^\prime (1-p^\prime)$. This simulation confirms the essential role
of the variance in the disordering process.

Note that in this section, we focus on the transformation which
drives the system from the SC state to the CPG state at zero
temperature, which is obtained by applying an adiabatic work
$W=\beta_c$. It should be distinguished from the transformation with
increasing temperature, where the pseudogap state is obtained at
$T_c$. Since the condensation energy is defined as the work needed
to destroy the correlations, an additional amount of energy has to
be furnished in order to destroy the coherent state (see Fig.
\ref{Fig_Transformations}) by heating, in agreement with the second
law of thermodynamics. While the CPG state is an incoherent state at
zero temperature, the pseudogap state at $T_c$ involves both pairon
and quasiparticle excitations
\cite{Jphys_Sacks2018,SolStatCom_Noat2021}. This dual aspect of the
thermal excitations is a determining factor in the shape of the
specific heat, for example, as shown in Ref.
\cite{SolStatCom_Noat2021}. Nevertheless, both the CPG and PG states
are essentially the same\,: a non-superconducting state of
uncorrelated pairons.


\subsection{Phenomenological expression for the energy of the
system}

In the BCS scenario the system gains energy by an ordering of Cooper
pairs in $k$-space, giving rise to the well-known spectral gap at
the Fermi level. However, in our mechanism, the energy gain is due
to pairon correlations in real space, on a length scale $l_c$. This
aspect can be understood by a phenomenological expression for the
energy $E(N_i)$ as a function of the local density of pairons
$N_i/N_0$ within a square box of side $l_c$, where $N_0$ is the
number of pairon lattice sites inside the box, i.e. $N_0$=$l_c^2$.

First, we note that the local exchange energy in the box, $J_{loc}$,
is reduced compared to $J_{eff}$ due
to the presence of $N_i$ pairons. It reads:
\begin{equation}
J_{loc}(N_i)=J_{eff}\,(1-{N_i}/{N_0})
\label{Equa_Jloc}
\end{equation}

As in the previous subsection, the total energy is found by
the statistical average over configurations.
In a given configuration we write $E(N_i)$ as the sum of 2 terms:
\begin{equation}
E(N_i)=-J_{loc}(N_i)- J_{loc}\,(N_i)\frac{N_i}{N_0}
\end{equation}
where the first term is the pairon self-energy (equivalent to the magnetic energy)
and the second term is the correlation energy.
Taking the statistical average of $E(N_i)$, we then obtain:
\begin{equation}
\left\langle E(N_i) \right\rangle=-\Delta_p-\beta_c+J_{eff}\frac{\sigma^2}{N_0^2}
\label{Equa_Etot}
\end{equation}
with
\begin{equation}
\Delta_p=J_{eff}\left(1-\frac{\left\langle N_i
\right\rangle}{N_0}\right) = J_{eff}\,(1-p') \label{Equa_Gap}
\end{equation}
and
\begin{equation}
\beta_c=J_{eff}\,\frac{\left\langle N_i \right\rangle}{N_0}\,
\left(1-\frac{\left\langle N_i \right\rangle}{N_0}\right) =
J_{eff}\,p'\,(1-p')
\end{equation}

\begin{figure}[t]
\includegraphics[width=9 cm]{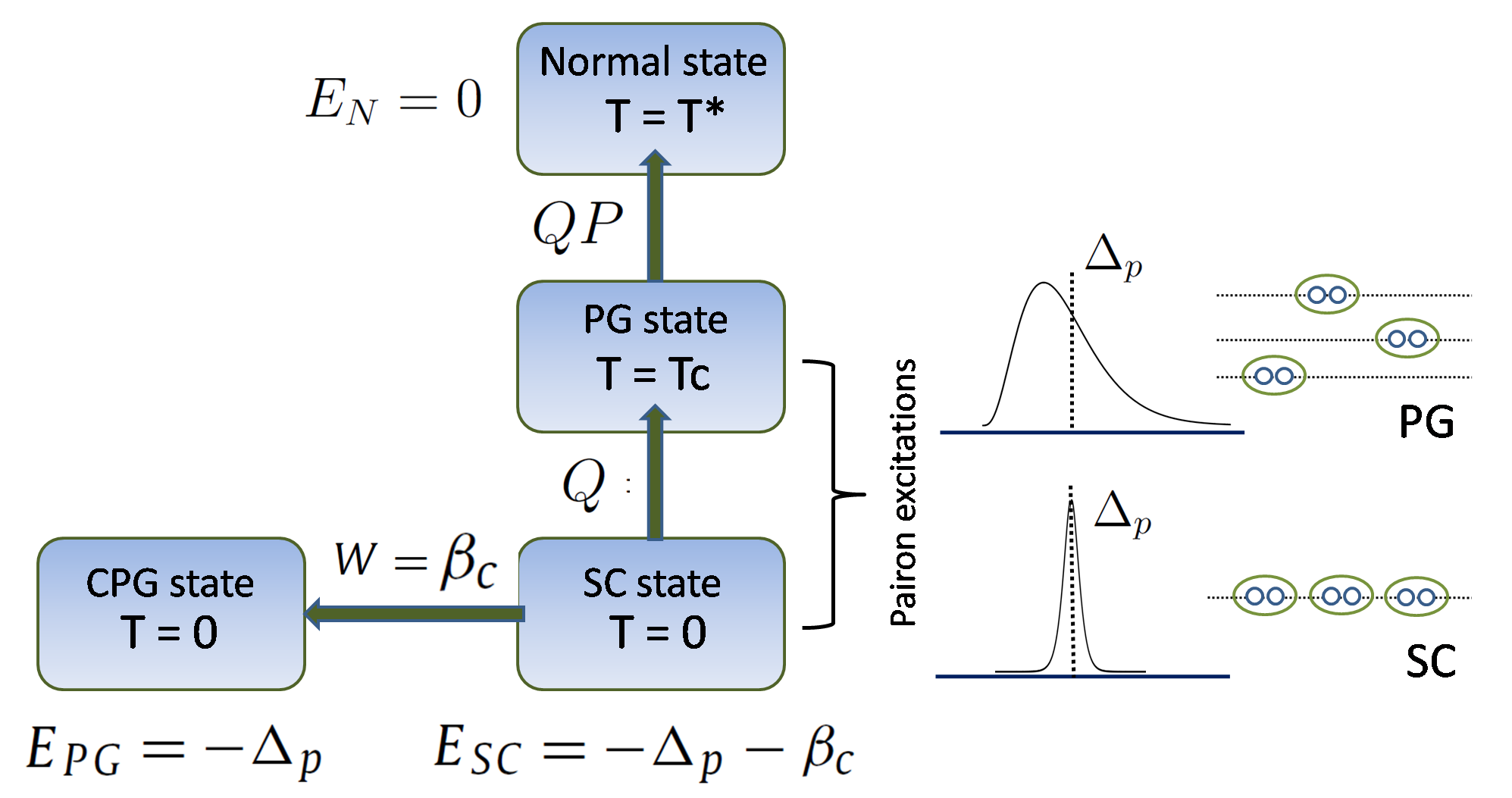}
\caption{(Color online) Left panel: Illustration of the two
different transformations: the first path at $T=0$ leads to the
disordered Cooper pair glass (CPG), an incoherent non SC state, by
applying an adiabatic work $W = \beta_c$ to the ordered SC state.
The second path, applying a thermal source, leads to the pseudogap
state by increasing the temperature, i.e. by furnishing an amount of
energy $Q$. Above $T_c$, the `normal state' is reached by
quasiparticle excitations (QP). Right panel: schematic distribution
of pairon energies in the SC state, where the distribution is
sharply peaked around the spectral gap $\Delta_p$. In the pseudogap
state, the energy distribution is very broad with a mean value
$\Delta_p$ and a characteristic width $2\,\sigma_v$.}
\label{Fig_Transformations}
\end{figure}


The third term in Eq. \ref{Equa_Etot} is a fluctuation term,
proportional to the variance $\sigma^2$ defined in Eq.
\ref{Equa_Variance}. With $N_0=l_c^2$, this third term takes the
form:
\begin{equation}
E_f =\frac{\beta_c}{l_c^2}
\label{Equa_Ef_lc}
\end{equation}
In the SC state where $l_c \rightarrow \infty$, $E_f=0$, so that:
\begin{equation}
E_{SC}=-\Delta_p-\beta_c
\label{Equa_Ener_SC}
\end{equation}
while in the CPG state where pairon correlations are absent, $l_c
\rightarrow 1$, $E_f=\beta_c$, so that
\begin{equation}
E_{CPG} =-\Delta_p
\end{equation}
Since the condensation energy is by definition the energy difference
between the CPG and the SC states, we find :
\begin{equation}
\Delta E=E_{CPG}-E_{SC}=\beta_c=J_{eff}\, p^\prime (1-p^\prime)
\label{Equa_diff_EPGSC}
\end{equation}

To describe intermediate states, it is useful to introduce an
effective correlation energy $\beta_{eff}$. Taking Eq.
\ref{Equa_Ef_lc} into account, Eq. \ref{Equa_Etot} can be written:
\begin{equation}
\left\langle E(N_i) \right\rangle=-\Delta_p-\beta_{eff}
\label{Equa_Eni_Beta_eff}
\end{equation}
with
\begin{equation}
\beta_{eff}=\beta_c \left(1-1/l_c^2 \right)
\label{Equa_Beta_eff}
\end{equation}
In such a way, the effective correlation energy $\beta_{eff}$ varies
from 1 (SC state) to 0 (CPG state) and can thus be considered as the
order parameter between the ordered and disordered states. Since
applying virtual work at zero temperature (by an applied magnetic
field) or thermally (by contact with a heat bath) the condensate
number density $N_{oc}(\lambda)$ will decrease monotonically (where
$\lambda$ is the relevant intensive variable) until reaching a
critical value $\lambda_c$ , i.e. the phase transition. Indeed, in
our previous work, we have shown that $N_{oc}(T)$ behaves as the
order parameter wherein the distribution of pairon excited states
$N_{ex}(T) = 1-N_{oc}(T)$ is governed by Bose-Einstein statistics
\cite{SciTech_Sacks2015} (the role of quasiparticle excitations was
given a full treatment in
\cite{Jphys_Sacks2018,SolStatCom_Noat2021}).

\begin{figure}
\includegraphics[width=8. cm]{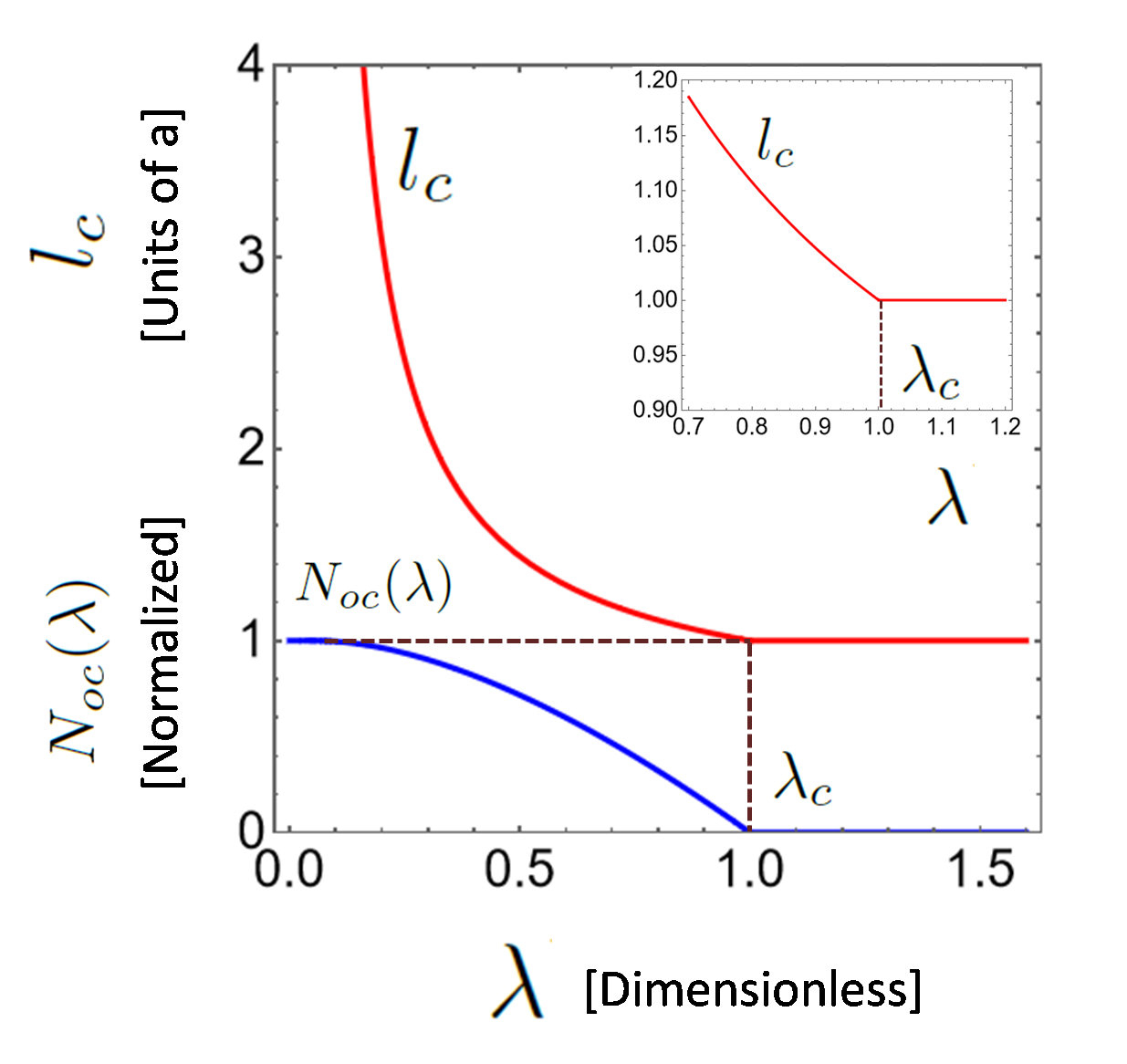}
\caption{(Color online) Plot of both the normalized condensate
density $N_{oc}(\lambda)$ and the correlation length $l_c$ (in units
of the pairon lattice $a$) as a function of the intensive variable
$\lambda$. The latter can represent either temperature or an applied
magnetic field which will destroy superconducting phase coherence
when $\lambda$ increases, reducing $N_{oc}$. Insert: zoom to show
the discontinuous derivative of $l_c(\lambda)$ at the critical
point. } \label{Fig_correlation length}
\end{figure}

The following general formula can then be proposed:
\begin{equation}
E_f= \beta_c\left[ 1-N_{oc}(\lambda) \right]
\label{Equa_Ef_Noc_lambda}
\end{equation}
Taking Eq. \ref{Equa_Ef_lc} into account, we find the relation
between the pairon correlation length $l_c$ and $N_{oc}$:
\begin{equation}
l_c(\lambda)=\frac{1}{\sqrt{\left(1-N_{oc}(\lambda)\right)}}
\label{Equa_lc}
\end{equation}
This equation represents a new critical length describing the
transition with a simple critical exponent $-\frac{1}{2}$. Moreover,
$l_c$ is interpreted as the correlation length of excited pairons in
the presence of the condensate. As shown in
Fig.\,\ref{Fig_correlation length}, $l_c$ diverges for
$\lambda\rightarrow 0$ ($N_{oc}(\lambda)\rightarrow 1$) and is equal
to 1, its minimal value, when $\lambda=\lambda_c$
($N_{oc}(\lambda_c)=0$) . As the inset of the Fig.
\ref{Fig_correlation length} shows, the derivative of $l_c$ is
discontinuous at the critical point.

Finally, the result for the condensation energy derived in this
work, $\Delta E = \beta_c$, captures a novel mechanism of the
transition. In the model, the condensate is a many-body quantum state of pairons
governed by the single energy scale $J_{eff}$ and the pairon-pairon
interaction distance $d_0$. The former replaces the traditional
phonon energy scale of the BCS theory \cite{PR_BCS1957}, while the
latter is the new length scale of the mean-field interaction in real
space. Once the condensate is destroyed by fluctuations, in the CPG state pairons
behave classically and are governed by the binomial law with
variance $\sim p^\prime (1-p^\prime)$.

The unconventional nature of the transition is striking: if this
were conventional SC, the condensation energy should be proportional
to the spectral gap at the Fermi level, $\Delta_p$. However, in this
approach, the condensation energy, $\beta_c \propto T_c$, is interpreted as the
correlation energy between pairons, which is maximum in the SC state
and vanishing in the CPG state. The fact that $\Delta_p$ is {\it
not} the order parameter remains one of the most intriguing aspects
of cuprate SC. Indeed, the average spectral gap at the Fermi level
is {\it invariant} from the ordered state to the CPG state (see Fig.
\ref{Fig_Transformations}, right panel).


\begin{figure}
\includegraphics[width=8.0 cm]{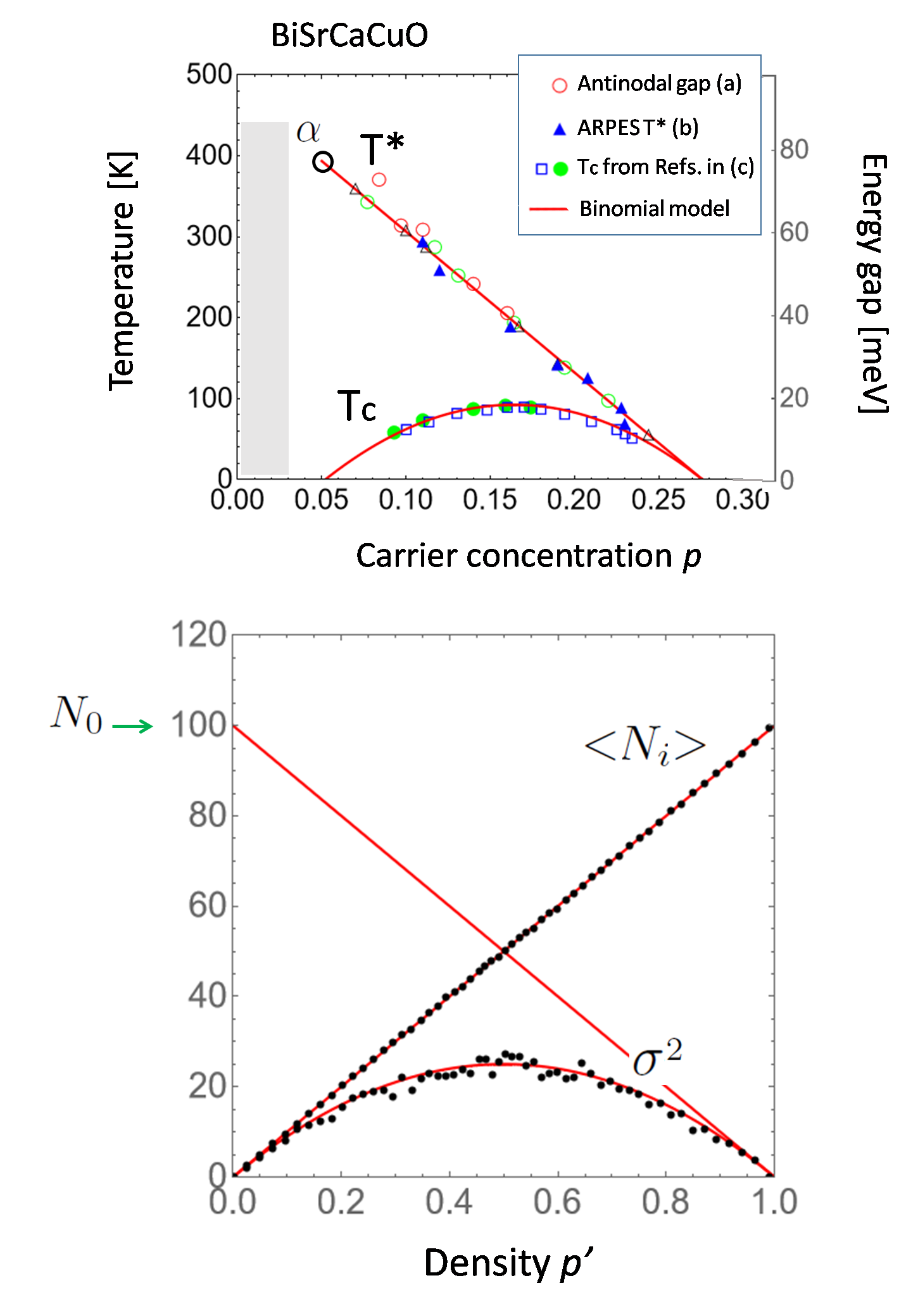}
\caption{(Color online) Upper panel: Theory compared to experiments:
(a) Antinodal gap (in meV, right hand scale) (b) pseudogap
temperature $T^*$, both measured by ARPES \cite{Nat_Hashimoto2014,
PNAS_Vishik2012} (c) experimental superconducting temperature $T_c$
(from Ref. \cite{PhysicaC_Tallon2001}), compared to the binomial law
(red lines) with two adjustable parameters $\alpha$ and $\alpha^\prime$. From the
fit of the data we find the values $\alpha=390$\,K and $\alpha^\prime=380$\,K. Lower panel:
Numerical simulation\,: Average number $<N_i>$ and variance
$\sigma_{N_i}^2$ as a function of reduced doping $p'$. }
\label{Fig_Simul_Exp}
\end{figure}

\begin{figure}
\includegraphics[width=9. cm]{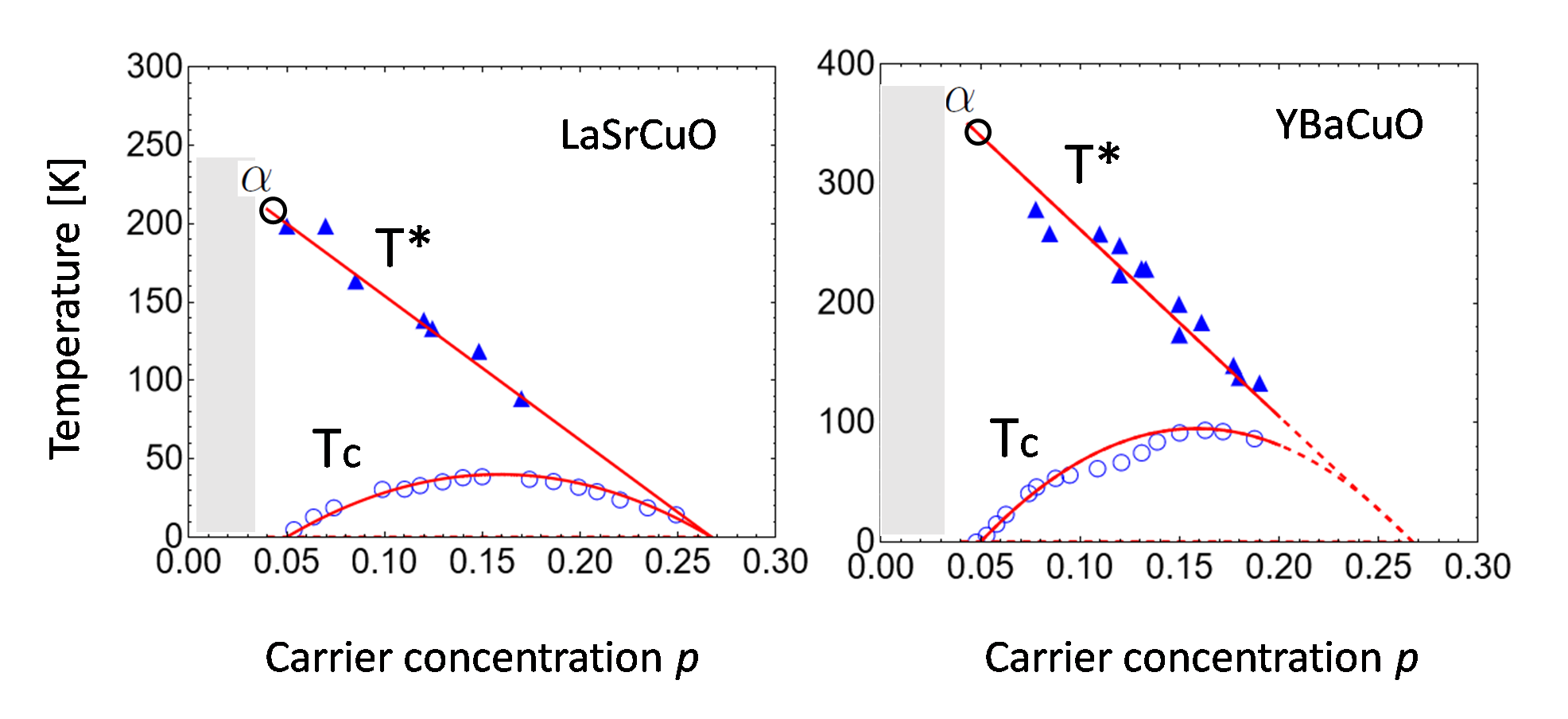}
\caption{(Color online)  Theory compared to experiments. Left panel:
critial temperature taken from Ref. \cite{PRL_Ando2004} and
pseudogap temperature $T^*$ for LSCO measured by Nernst effect (blue
triangles) taken from \cite{{PRB_Cyr-Choiniere2018}}. and comparison
with the model (red curves) with the two adjustable parameters
$\alpha=200$K and $\alpha^\prime=250$K. Right panel: critical
temperature taken from Ref. \cite{PRB_Liang2006} and pseudogap
temperature $T^*$ for YBCO measured by Nernst effect (blue
triangles) taken from \cite{{PRB_Cyr-Choiniere2018}} and comparison
with the model (red curves) with $\alpha=340$K and
$\alpha^\prime=380$K . } \label{Fig_LSCO_YBCO}
\end{figure}

\subsection{Comparison with experiments for different materials}

As we have discussed, the pseudogap temperature $T^*$ decreases with
hole concentration while the critical temperature is nearly a
perfect dome. We then hypothesize that both $T_c$ and $T^*$ emerge
from the proposed energy functional and statistical approach
described above. This is a continuation of our previous work
\cite{PhysLettA_Noat2022} wherein the superconducting state and the
pseudogap state are not independent phenomena. On the contrary, they
are intimately coupled, so that $T_c$ and $T^*$ follow the
relations:
\begin{eqnarray}
&&T^*(p^\prime)=\alpha\, (1-p^\prime)  \nonumber\\
&& T_c(p^\prime)=\alpha^\prime\, p^\prime(1-p^\prime)
\label{Equa_Tstar_Tc}
\end{eqnarray}
where $\alpha $ and  $\alpha^\prime$ depend on the the {\it same} energy scale parameter $J_{eff}$.

We then compare the robustness of the model for BSCCO$_{2212}$ to
the experimental values of the gap and $T^*$ measured by
photoemission (Fig. \ref{Fig_Simul_Exp}) as well as the $T_c$ dome.
The agreement is remarkable for a value of the two parameters
$\alpha=$390\,K and $\alpha^\prime=$380\,K. The values found for
$\alpha$ and $\alpha^\prime$ are very close. Indeed, we found in our
previous article \cite{PhysLettA_Noat2022} that for BSCCO$_{2212}$
$\alpha$ and $\alpha^\prime$ can be taken as equal to the
experimental uncertainty. (Note that $\alpha^\prime=4\times
T_c^{max}$.)

From the $T_c$
value for a certain doping, we can deduce the expected value
for the pseudogap temperature, proving their intimate link.
\\We also examine the case of LSCO and YBCO (see Fig. \ref{Fig_LSCO_YBCO}). Taking the values of the pseudogap temperature
found from Nernst experiments \cite{{PRB_Cyr-Choiniere2018}}, we
obtain the values for the two adjustable parameters. For LSCO, we
find, $\alpha=$200\,K and $\alpha^\prime=$160\,K. For YBCO we find
$\alpha=$340\,K and $\alpha^\prime=$380\,K. The results are
summarized in Table \ref{Table_param}.
\begin{table}[]
\begin{tabular}{llll}
            & $\alpha$(K)  & $\alpha^\prime$(K) &  $\alpha/\alpha^\prime$\\ \hline
BSCCO$_{2212}$ & 390 & 380 &    1.02                                  \\ \hline
LSCO        & 200 & 160  &    1.25                                 \\ \hline
YBCO        & 340 & 380 &    0.89  \\ \hline
\end{tabular}
\caption{(Color online)  Comparison of the values for the two
adjustable parameters $\alpha$ and $\alpha^\prime$ found from the
fits for the three different cuprates, BSCCO$_{2212}$, LSCO and
YBCO.} \label{Table_param}
\end{table}

We find that YBCO is quite similar to BSCCO$_{2212}$ although $T^*$
is slightly smaller than expected. On the other hand, $T^*$ for LSCO
is significantly smaller and $\alpha/\alpha^\prime\simeq 1.25$. The
differences in the ratio are most likely due to the structure of the
different materials and the fact that LSCO has a single CuO plane
and weak interlayer coupling compared to BSCCO$_{2212}$.

Since $\Delta_p=J_{eff} (1-p^\prime) \simeq 2.2 k_B
T^*$ from ARPES and tunneling (see Fig. \ref{Fig_Simul_Exp}), we get
numerically $J_{eff} = 2.2\, k_B \alpha \simeq$ 74 meV BSCCO$_{2212}$. This yields
the value of the spectral gap at optimal doping ($p^\prime=0.5$, $p=0.165$ )\,: $\Delta_p
\simeq 74/2 = 37$\,meV. Using the expression for $\beta_c$,
the SC condensation energy at optimal doping is $\beta_c \simeq 74/4
= 18.5$\,meV, a very reasonable value.

Finally, the Ginzburg-Landau coherence length associated with $\beta_c$ is written:
\begin{equation}
\xi=\sqrt{\frac{\hbar^2}{2m\beta_c}}
\label{Equa_Xi_GL}
\end{equation}
Using the numerical value of $\beta_c$, we obtain $\xi\sim 1.4$ nm,
which is typically the length scale of the vortex core in cuprates
\cite{PRL_Pan2000}.

\subsection{Statistics of the Voronoi cell areas}

An equally important geometrical property of random points on a
square lattice is the Voronoi cell area distribution. Such a
mathematical analysis is used to study real physical systems such as
confined charged beads in 2D \cite{EPJB_StJean2004}. In this
paragraph we connect their statistics to the physical properties of
pairons, which extends the previous analysis of density
fluctuations. The distribution of the Voronoi cell areas $A_i$
surrounding each pairon is calculated as a function of
concentration.

A typical result is shown in Fig.\ref{Fig_Statistics}, lower panel. We find an
asymmetric distribution in agreement with previous mathematical
studies \cite{Form_Tanemura2003,PhysicaA_Szabo2007}. As a
consequence, the mean value is larger than the maximum of the
distribution which is the most probable value. Qualitatively, we see
that the distribution is much broader than the binomial distribution
and extends to very large areas. We find that a relatively simple
skewed distribution function matches the numerical simulation
satisfactorily at all concentrations. It reads:
\begin{equation}
P_v(A_i)=(A_i-A_c)^b\exp(-A_i/c) \label{Equa_Distrib_Voro}
\end{equation}
where $b$, $c$ are free fitting parameters, and $A_c$ is the minimum area of a pairon.

For a given density, the average of $A_i$ and the average of $N_i$
are linked by the constraint:
\begin{equation}
\frac{\left\langle A_i\right\rangle}{A_0}=\frac{\left\langle N_i\right\rangle}{N_0}=p^\prime
\label{Equa_Moy_Ni_Ai}
\end{equation}
where $A_0$ is the sampling area of the density.

Following our previous work \cite{EPL_Sacks2017}, we assume that
each pairon binding energy $\Delta_i$ is determined by the surface
area of the Voronoi cell:
\begin{equation}
\Delta_i = J_{eff}\, \frac{A_i-A_c}{\left\langle A_i\right\rangle}
\label{Equa_Delta_i}
\end{equation}
This hypothesis means that the binding energy of a given pairon is
proportional to the number of coherent spins in a given Voronoi
cell, on the scale of $\xi_{AF}$.Since $\xi_{AF}(p)\approx
a_0\sqrt{(2/p)}$ \cite{PRB_Birgeneau1988}, where the additional
$\sqrt{2}$ comes from the formation of hole pairs below $T^*$,
$\xi_{AF}(p)$ varies from $\xi_{AF}\sim 6.02\,a_0$ for $p=0.05$ to
$\xi_{AF}\sim 2.72\,a_0$ for $p=0.27$ along the superconducting
dome.


Taking the average value of $\Delta_i$, one obtains:
\begin{equation}
\Delta_p=\left\langle \Delta_i\right\rangle=J_{eff}(1-p^\prime)
\end{equation}
so that Eq.\ref{Equa_Gap} is recovered. We find that the mean binding
energy $\Delta_p(p^\prime)$ decreases linearly with $p^\prime$ as
expressed in the energy functional and verified in the density
fluctuation simulation. Note that this binding energy corresponds to
the spectral gap at the Fermi level as measured by ARPES and
tunneling \cite{SciTech_Sacks2015}.

\begin{figure}
\includegraphics[width=7.0 cm]{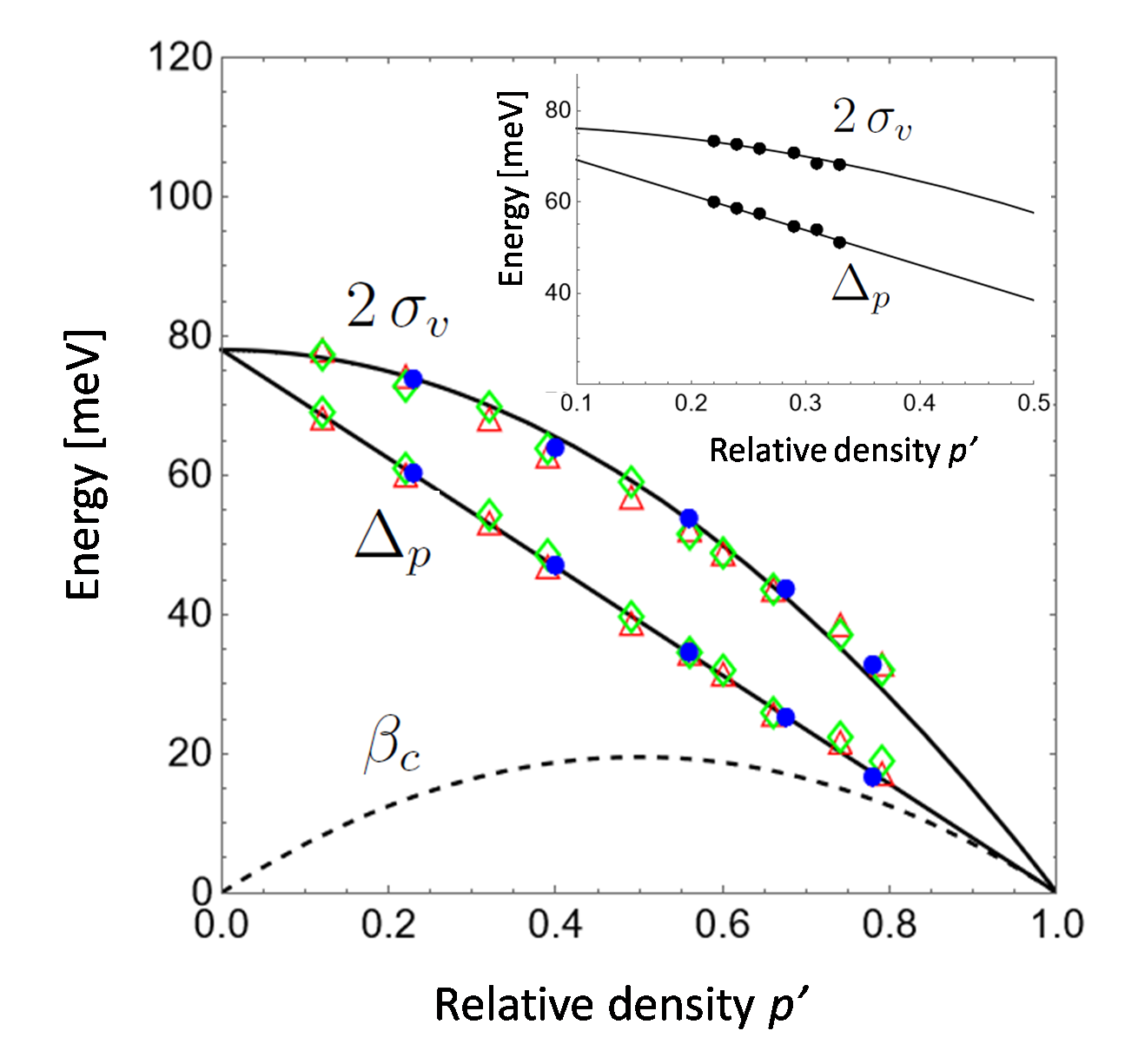}
\caption{(Color online) Parameters deduced from the statistics of
the Voronoi cells, the mean value and the variance, converted in
energy units (see text). Triangles and diamonds use two different
sample sizes ($N_T$= 40$\times$40 and $N_T$ = 50$\times$50,
respectively), while the blue dots are extracted using the
distribution, Eq.\ref{Equa_Distrib_Voro}, for the best fit. The
magnitude of the jitter is roughly the uncertainty. The black lines
correspond to the expected mean value binding energy $\Delta_p$
(lower straight line) and to the total energy $\Delta_0$ (upper
line), respectively. Inset: sampling points on the underdoped side
$p'<.5$.} \label{Fig_Param_Voronoi}
\end{figure}

We then calculate the variance of the Voronoi cell areas
$\sigma_{A}^2=\left\langle A_i^2\right\rangle-\left\langle
A_i\right\rangle^2$. For comparaison to physical quantities, we
convert $\sigma_{A}$ to the same energy scale using the relation :
$$\sigma_v=J_{eff}\frac{\sigma_{A}}{\left\langle A_i\right\rangle}$$
Similarly, the average binding energy $\Delta_p(p^\prime)$ is calculated
from the Voronoi statistics using equation \ref{Equa_Delta_i}. The calculations
were then done systematically as a function of doping $p^\prime$
throughout the phase diagram, independently of the density
fluctuation approach. The combined results for $\Delta_p(p^\prime)$ and
$2\sigma_v(p^\prime)$ are summarized in the plots of Fig.
\ref{Fig_Param_Voronoi}.

Significantly, we find that $\sigma_v$ has an important physical
meaning. From the calculated Voronoi distribution in the disordered
CPG state, $2\,\sigma_v$ is found to be very close to the sum
$\Delta_0=\Delta_p+\beta_c$. More precisely, as can be seen in
Fig.\,\ref{Fig_Param_Voronoi}\,:
\begin{equation}
2\,\sigma_v \simeq \Delta_0
\end{equation}
for a wide range of $p^\prime$. (A typical fit is\,: $2\,\sigma_v =
\Delta_0/1.04$, which is quite satisfactory.) In fact the quantity
$\Delta_0$ appears naturally in the energy functional as described
previously (see Eq. \ref{Equa_Ener_SC}); it corresponds to the total
energy of the system in the SC state\,:
\begin{equation}
E_{SC}=-\Delta_p-\beta_c=-\Delta_0
\label{Equa_E_SC}
\end{equation}
i.e. the  sum of the binding energy and the correlation energy.
Thus, we expect $\Delta_0$ to be revealed in experiments probing the
cuprate superconducting state.

In summary, the physical quantities can be understood from Fig. \ref{Fig_Transformations} showing
that the CPG is obtained by the adiabatic transformation at zero
temperature by applying a work $W=\beta_c$ to the SC state. The
examination of Fig.\,\ref{Fig_Param_Voronoi} reveals unequivocally
that the fundamental SC/PG parameters also appear in the Voronoi
distribution. Remarkably, the statistics of the cell areas contains
the same information as the density fluctuations.

Another key point is revealed by the distribution of Voronoi cells
of the CPG state\,: in a given configuration there is a relatively
wide distribution of areas (the $A_i$) over a width $2\,\sigma_{A}$
(see Fig.\ref{Fig_Statistics}, lower right panel). This implies that
in the CPG state, there exists an equivalent distribution of pairon
energies $\Delta_i$ with a width equal to the total energy of the
system, $\Delta_0$. Therefore, a key characteristic of the pseudogap
state is the energy distribution of `pre-formed pairs' of width $2\sigma_v
= \Delta_0$.

In our previous work, we extended this idea of pair
energy distribution to the pair {\it excited states} in the
calculation of the ARPES quasiparticle spectral function \cite{Jphys_Sacks2018} and
the tunneling density of states \cite{SciTech_Sacks2015}. For the excited states, we
assumed a Lorentzian distribution\,:
\begin{equation}
P_{0}(\Delta_i) =
\frac{\sigma_0^2}{(\Delta_i-\Delta_0)^2+\sigma_0^2} \label{Equa_Lor}
\end{equation}
whose physical interpretation is the energy spread over a width
$2\,\sigma_0$ of a system of pairons of total energy $\Delta_0$. The
conclusion of this subsection is that $\sigma_0$ can now be
identified as $\sigma_v$, and thus it is no longer an independent
parameter of the problem.

In conclusion, the fundamental parameters characterizing the SC
state and the PG state, namely the condensation energy $\beta_c$ and
the pseudogap $\Delta_p$, both emerge from the statistical
properties of the disordered state. The only free parameter of the
problem left is therefore the exchange energy, $J_{eff}$.

\section{Quasiparticle density of states}

\begin{figure}
\includegraphics[width=8.4 cm]{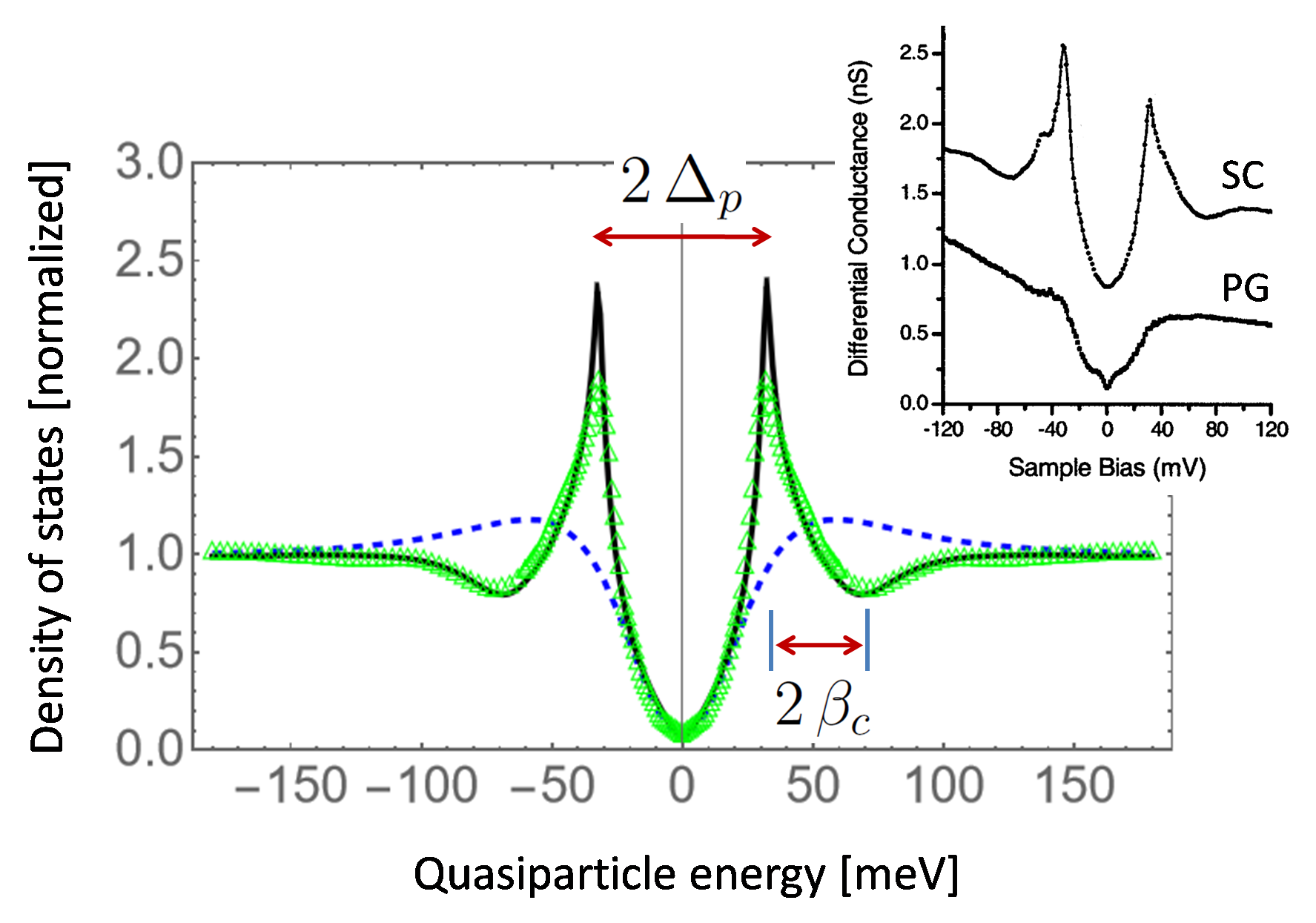}
\caption{(Color online) Experimental quasiparticle spectrum in the
SC state (green triangles) taken from Ref. \cite{PRL_Pan2000} (note
that the experimental spectrum for $E>0$ was considered and
normalized for clarity) and fit (black line) using the pairon model
and the gap function, Eq. \,\ref{Equa_gapfun}. The fit of the
peak-dip-hump structure is accurate using the values\,: $J_{eff} =
74$ meV and $p =0.178$. The important parameters from the
statistical model, $\Delta_p$ and $\beta_c$, are indicated by the
arrows. Dashed line: quasiparticle spectrum calculated using the
pairon energies deduced from the statistics of Voronoi cells in the
disordered state. Inset: Experimental tunneling spectrum in the SC
state and in the vortex core \cite{PRL_Pan2000}.} \label{Fig_QP
spectrum}
\end{figure}

In this final section we show that the fundamental parameters of the
model are seen in the low-temperature tunneling spectra. In the
previous section, we have seen that the pairon binding energies
$\Delta_i$, calculated from the  Voronoi cell areas in the
disordered CPG state, represent the available pairon states once SC
coherence is destroyed (see Fig. \ref{Fig_Transformations}). Such a
state, obtained in a {\it gedanken} experiment by applying an adiabatic
work $W=\beta_c$ to the system, is realized in a {\it bona fide} STM/STS
experiment in a vortex core, where SC coherence is broken, Fig. \ref{Fig_QP spectrum}. This is
commonly referred to as the low-temperature pseudogap in the
quasiparticle density of sates (DOS).

The theoretical quasiparticle spectrum in the CPG state can be obtained from
the pairon binding energies $\Delta_i$ deduced from the previously
determined Voronoi distribution $P_{v}(\Delta_i)$. The latter
distribution has a width $2\,\sigma_v$ and average gap $\Delta_p$,
as previously described. These two parameters are known for any
value of the carrier concentration, $p$, with the choice $J_{eff}
\simeq 74$\,meV. We use the simple formula\,:
\begin{equation}
N_{CPG}(E)=N_n\ \sum_i\ P_{v}(\Delta_i)\,\frac{\left |E\right |} {\sqrt{E^2-\Delta_i^2}}
\label{Equa_DOS_CPG}
\end{equation}
where $N_n$ is a constant, and the integrand is the quasiparticle
DOS associated with the pairon energy level $\Delta_i$. The
resulting PG-like DOS is plotted in Fig.\,\ref{Fig_QP spectrum}
(dashed line) and compared to the experimental one found in the
vortex core (inset of Fig.\,\ref{Fig_QP spectrum}) obtained by Pan
et al. \cite{PRL_Pan2000}.

The quasiparticle DOS in the SC state is more
involved to calculate and was given full treatments in
\cite{PRB_Sacks2006,SciTech_Sacks2015,Jphys_Sacks2017}. Indeed, in the coherent
state, the self-consistent $d$-wave gap function was shown to be\,:
\begin{equation}
\Delta_{\vec k}(E_{\vec k}) = \Delta_p\,\left(\cos(k_x a_0) -
\cos(k_y a_0)\right)\,\left(1 - 2\,\beta_c\, P_{0}(E_{\vec k}) \right)
\label{Equa_gapfun}
\end{equation}
where $E_{\vec k}=\sqrt{\varepsilon_k^2+\Delta_{\vec k}(E_{\vec
k})^2}$ is the quasiparticle energy, with $\varepsilon_k$ the
kinetic energy with respect to the Fermi energy. The above gap
function has a complex pole at the value $\tilde E_{\vec k} =
\Delta_0 + i\,\sigma_0$ \cite{EPL_Cren2000,PRB_Sacks2006} giving
rise to the characteristic sharp dip structure in the quasiparticle
DOS at the position $E = \Delta_p + 2\,\beta_c$. A similar pole
gives the dip for negative energies.

The experimental quasiparticle spectrum is shown in
Fig.\,\ref{Fig_QP spectrum}, green triangles, compared to the
theoretical curve using the gap function of Eq. \ref{Equa_gapfun}, solid
black line. It exhibits the very sharp, but characteristically wide,
coherence peaks, as well as the aforementioned dip-hump structure.
We recall that, in the model, the two basic parameters $\Delta_p$
and $\beta_c$ are known as a function of the carrier density, $p$ and
the effective exchange interaction $J_{eff}$. Since the latter is
fixed, the fit in Fig.\,\ref{Fig_QP spectrum} can be done with
essentially one free parameter, in this instance $p=.178$. In
practice, we do include some quasiparticle damping (Dynes $\Gamma$
broadening \cite{PRL_Dynes1978}) to achieve the best fit \cite{PRB_Sacks2006}.

In the disordered CPG state, the shape of the QP spectrum is very
different: the coherence peaks are almost completely smeared out, as
found in the real experimental spectrum (Fig. \ref{Fig_QP spectrum},
right inset). Note that a very similar spectrum is also observed in
a highly disordered region \cite{PRL_Cren2000, PRB_Howald2001}. The
QP spectrum inside a vortex core thus reflects the CPG state where
pairs occupy randomly the available pair states of the distribution,
without global phase coherence.

In conclusion, the essential parameters of the model, the low-
temperature gap $\Delta_p$ and the condensation energy
$\beta_c$, can be directly extracted from tunneling experiments,
either in the SC state, or in the CPG state, realized outside or within the vortex core, respectively.

\section{Conclusion}

We have studied the statistics of pairons randomly distributed on a
square lattice. We have shown that the fundamental aspects of the
phase diagram, the $T_c$ dome, the pseudogap temperature $T^*$,
emerge from the statistical analysis of the disordered state of
pairons.

The SC transition appears as a disorder to order transition,
resulting in a correlated state of pairons in real space, the
condensate. A novel expression for the pairon critical correlation
length is derived as a function of the consensate density. The
coherence energy is the adiabatic work needed to drive the system
from the ordered SC state to an incoherent state of pairons, the
Cooper pair glass state, to which it is intimately connected. It is
shown that SC coherence is achieved through correlations of pairons
following a binomial law, explaining the superconducting dome.

The study of two quantum entities, the simplons and pairons, and
their condensation, allows to gain insight on the temperature scales
of the phase diagram, $T_{max}$, $T^*$, and $T_c$ and why they can
be unified amongst the cuprates. In particular, the statistical
analysis of pairon states on a square lattice proves that the key
energy parameters, the spectral gap $\Delta_p$ and the condensation
energy $\beta_c$, are proportional to only one energy scale
$J_{eff}$.

The inevitable conclusion is that the superconducting and the
pseudogap states emerge from the same interaction. This interaction
is no doubt the local antiferromagnetic interaction, which accounts
for the binding energy of pairons, in addition to the correlation
energy between them. It appears that the pairon condensate is the
ground state of a system where superconductivity and magnetism would
otherwise be antagonistic. The total energy of the ground state is
shown to be $E_{SC} = -\Delta_p - \beta_c$, i.e. the sum of the
pairon self-energy, and the correlation energy, respectively.
Contrary to conventional superconductivity, it is the latter energy
that is tied to the SC coherence.

Finally, the essential parameters of the energy states, together
with the numerical value of $J_{eff}$, allows to predict
quantitatively tunneling spectroscopy experiments at low
temperature.
\vskip 5 mm
{\bf Acknowledgements}\\
We gratefully acknowledge Takao Morinari, Jeff Tallon and Atsushi
Fujimori for interesting and useful discussions.


\vskip 2 mm


\begin{thebibliography}{99}

\bibitem{ZPhys_Bednorz1986} J. G. Bednorz, K. A. M\"{u}ller, Possible high $T_c$ superconductivity in the Ba--La--Cu--O system , Zeitschrift f\"{u}r Physik B Condensed Matter {\bf 64}, 189 (1986).

\bibitem{FrontPhys_Hott2004} Roland Hott, Reinhold Kleiner, Thomas Wolf and Gertrud Zwicknag, Superconducting Materials -- A Topical Overview, in Frontiers in Superconducting Materials, Ed. Anant V. Narlikar, Springer Verlag, Berlin, pp 1-69 (2004).

\bibitem{JPhysChem_Tahir2010} Jamil Tahir-Kheli and William A. Goddard III, Universal Properties of Cuprate Superconductors: Tc Phase Diagram, Room-Temperature Thermopower, Neutron Spin Resonance, and STM Incommensurability Explained in Terms of Chiral Plaquette Pairing, J. Phys. Chem. Lett. {\bf 1}, 1290--1295 (2010).

\bibitem{PhysLettA_Noat2022} Yves Noat, Alain Mauger, William Sacks. Superconductivity in cuprates governed by topological constraints. Physics Letters A {\bf 444}, 128227 (2022).

\bibitem{RepProgPhys_Hufner2008} S. H\"ufner, M. A. Hossain, A Damascelli, and G. A. Sawatzky,Two gaps make a high-temperature superconductor?, Rep. Prog. Phys., {\bf 71}, 062501 (2008).

\bibitem{PRB_Keimer1992} B. Keimer, N. Belk, R. J. Birgeneau, A. Cassanho, C. Y. Chen, M. Greven, M. A. Kastner, A. Aharony, Y. Endoh, R. W. Erwin, and G. Shirane, Magnetic excitations in pure, lightly doped, and weakly metallic La$_2$CuO$_4$, Phys. Rev. B {\bf 46}, 14034 (1992).

\bibitem{PhysicaC_Presland1991} M.R. Presland, J.L. Tallon, R.G. Buckley, R.S. Liu and N.E. Flower,
General trends in oxygen stoichiometry effects on Tc in Bi and Tl
superconductors, Physica C, PhysicaC {\bf 176}, 95--105  (1991).

\bibitem{PRL_Torrance_1988}
J. B. Torrance, Y. Tokura, A. I. Nazzal, A. Bezinge, T. C. Huang, and S. S. P. Parkin
Anomalous Disappearance of High-T$_c$ Superconductivity at High Hole Concentration in Metallic La$_{2-x}$Sr$_x$Cu0$_4$,
Phys. Rev. Lett. {\bf 61}, 1127 (1988).

\bibitem{PRB_Takagi1989} H. Takagi, T. Ido, S. Ishibashi, M. Uota, S. Uchida, and Y. Tokura, Superconductor-to-nonsuperconductor transition in (La$_{1-x}$Sr$_x$)$_2$CuO$_4$ as investigated by transport and magnetic measurements, Phys. Rev. B {\bf 40}, 2254 (1989).

\bibitem{PRL_Ando2004} Yoichi Ando, Seiki Komiya, Kouji Segawa, S. Ono, and Y. Kurita,
Electronic Phase Diagram of High-T$_c$ Cuprate Superconductors from a Mapping of the In-Plane Resistivity Curvature,
Phys. Rev. Lett. {\bf 93}, 267001 (2004).

\bibitem{PRB_Feng2012} Shiping Feng, Huaisong Zhao, and Zheyu Huang, Two gaps with one energy scale in cuprate superconductors, Phys. Rev. B {\bf  85}, 054509 (2012).

\bibitem{SUST_Marino2020} E. C. Marino, Reginaldo C. , Corr\'ea Jr, Lizardo H. C. M. Nunes,Van S\'ergio Alves, Superconducting and pseudogap transition temperatures in high-Tc cuprates and the Tc dependence on pressure,  Supercond. Sci. Technol. {\bf 33}, 035009 (2020).

\bibitem{PR_BCS1957} J. Bardeen, L. Cooper, J. Schrieffer, Theory of Superconductivity, Phys. Rev. {\bf 108} 1175 (1957).

\bibitem{PR_Cooper1956} Leon N. Cooper, Bound electron pairs in a degenerate Fermi gas, Physical Review {\bf 104}, 1189-1190 (1956).

\bibitem{PR_Giaever1962} I. Giaever, H. R. Hart, Jr., and K. Megerle, Tunneling into Superconductors at Temperatures below 1$^{\circ}$ K, Phys. Rev. {\bf 126}, 941 (1962).

\bibitem{PhysicaC_Bosovic2019} Bo\v{z}ovi\'{c}, J. Wu, X. He, A.T. Bollinger, What is really extraordinary in cuprate superconductors?, Physica C {\bf  558}, 30--37 (2019).

\bibitem{Revmod_Fisher2007} \O. Fischer, M. Kugler, I. Maggio-Aprile,
C. Berthod and C. Renner, Scanning tunneling spectroscopy of the
cuprates, Rev. Mod. Phys. {\bf 79}, 353 (2007).

\bibitem{PhysicaC_Loram1994}
J.W. Loram, K.A. Mirza, J.M. Wade, J.R. Cooper, W.Y. Liang
The electronic specific heat of cuprate superconductors
Physica C {\bf 235--240}, 134-137 (1994).

\bibitem{JPhysSocJp_Matsuzaki2004} Toshiaki Matsuzaki, Naoki Momono, Migaku Oda, and Masayuki Ido, Electronic Specific Heat of La$_{2-x}$Sr$_x$CuO$_4$: Pseudogap Formation and Reduction
of the Superconducting Condensation Energy, J. Phys. Soc. Jpn. {\bf 73}, 2232 (2004).

\bibitem{PRL_Wen2009} Hai--Hu Wen, Gang Mu, Huiqian Luo, Huan Yang, Lei Shan, Cong Ren, Peng Cheng, Jing Yan, and Lei Fang,Specific--Heat Measurement of a Residual Superconducting State in the Normal State of Underdoped Bi$_2$Sr$_{2-x}$La$_x$CuO$_{6+\delta}$ Cuprate Superconductors, Phys. Rev. Lett. {\bf 103}, 067002 (2009).

\bibitem{Nat_Xu2001} Z. A. Xu, N. P. Ong, Y. Wang, T. Kakeshita and S. Uchida,
Vortex-like excitations and the onset of superconducting phase
fluctuation in underdoped La$_{2-x}$Sr$_x$CuO$_4$, Nature {\bf 406},
486 - 488 (2000).

\bibitem{PRB_Wang2006} Yayu Wang, Lu Li, and N. P. Ong, Nernst effect in high-$T_c$ superconductors, Phys. Rev. B {\bf 73}, 024510 (2006).

\bibitem{PRB_Li2010} Lu Li, Yayu Wang, Seiki Komiya, Shimpei Ono, Yoichi Ando, G. D. Gu, and N. P. Ong, Diamagnetism and Cooper pairing above $T_c$ in cuprates, Phys. Rev. B {\bf 81}, 054510 (2010).

\bibitem{Nat_Zhou2019} Panpan Zhou, Liyang Chen, Yue Liu, Ilya Sochnikov, Anthony T. Bollinger, Myung-Geun Han, Yimei Zhu, Xi He, Ivan Bo\v{z}ovi\'{c} and Douglas Natelson
Electron pairing in the pseudogap state revealed by shot noise in copper oxide junctions, Nature {\bf 572}, 493--496 (2019).

\bibitem{PRL_Takagi1992}
H. Takagi, B. Batlogg, H. L. Kao, J. Kwo, R. J. Cava, J. J. Krajewski, and W. F. Peck, Jr.
Systematic evolution of temperature-dependent resistivity in La$_{2-x}$Sr$_x$CuO$_4$,
Phys. Rev. Lett. {\bf 69}, 2975 (1992).

\bibitem{PRL_Ito1993} T. Ito, K. Takenaka, and S. Uchida, Systematic deviation from T-linear behavior in the in-plane resistivity of YBa$_2$Cu$_3$O$_{7-y}$: Evidence for dominant spin scattering, Phys. Rev. Lett. {\bf 70}, 3995 (1993).

\bibitem{Rep_ProgPhys_Timusk1999} T. Timusk and B. Statt, The pseudogap in high--temperature superconductors:
An experimental survey. Rep. Prog. Phys. {\bf 62}, 61-122 (1999).

\bibitem{LowTempPhys_Kordyuk2015} A. A. Kordyuk, Pseudogap from ARPES experiment: Three gaps in cuprates and topological superconductivity, Low Temp. Phys.  {\bf 41}, 319 (2015).

\bibitem{PRL_renner1998_T} Ch. Renner,B. Revaz, J.-Y. Genoud, K. Kadowaki,and {{\O}}. Fischer, Pseudogap precursor of the superconducting gap in under- and overdoped Bi$_2$Sr$_2$CaCu$_2$O$_{8+\delta}$, Phys. Rev. Lett., {\bf 80} 149 (1998).

\bibitem{JphysSocJap_Sekine2016} R. Sekine, S. J. Denholme, A. Tsukada, S. Kawashima, M. Minematsu,T. Inose, S. Mikusu, K. Tokiwa, T. Watanabe, and N. Miyakawa, Characteristic features of the mode energy estimated from tunneling conductance on TlBa$_2$Ca$_2$Cu$_3$O$_{8.5+\delta}$, J. Phys. Soc. Jpn. {\bf 85}, 024702 (2016).

\bibitem{PRL_Miyakawa1999} N. Miyakawa, J. F. Zasadzinski, L. Ozyuzer, P. Guptasarma, D. G. Hinks, C. Kendziora, and K. E. Gray,
Predominantly Superconducting Origin of Large Energy Gaps in Underdoped
Bi$_2$Sr$_2$CaCu$_2$O$_{8+\delta}$ from Tunneling Spectroscopy, Phys. Rev. Lett. {\bf 83}, 1018 (1999).

\bibitem{JphysConfSer_Sugimoto2021} Akira Sugimoto, Hironori Ohtsubo, Kaito Matsumoto, Satoru Ishimitsu, Masatoshi
Iwano, Toshikazu Ekino, and A M Gabovic, Tunneling STM/STS and break-junction spectroscopy of the Pb-doped Bi2223 superconductor, J. Phys.: Conf. Ser. {\bf 1975}, 012005 (2021).

\bibitem{Revmod_Damascelli2003} Andrea Damascelli, Zahid Hussain, and Zhi-Xun Shen, Angle-resolved photoemission studies of the cuprate superconductors, Rev. Mod. Phys.  {\bf 75}, 473 (2003).

\bibitem{PNAS_Chatterjee2011} Utpal Chatterjee, Dingfei Ai, Junjing Zhao, Stephan Rosenkranz, Adam Kaminski, Helene Raffy, Zhizhong Li, Kazuo Kadowaki, Mohit Randeria, Michael R. Norman, and J. C. Campuzano, Electronic phase diagram of high-temperature copper oxide superconductors, PNAS  {\bf 108}, 9346-9349 (2011).

\bibitem{Natcom_Anzai2013} H. Anzai, A. Ino, M. Arita, H. Namatame, M. Taniguchi, M. Ishikado, K. Fujita, S. Ishida and S. Uchida, Relation between the nodal and antinodal gap and critical temperature in superconducting Bi2212, Nature Communications {\bf 4}, 1815 (2013).

\bibitem{JphysSocJp_Yoshida2012} Teppei Yoshida, Makoto Hashimoto, Inna M. Vishik, Zhi-Xun Shen, and Atsushi Fujimori,
Pseudogap, Superconducting Gap, and Fermi Arc in High-Tc Cuprates Revealed by Angle-Resolved Photoemission Spectroscopy,
J. Phys. Soc. Jpn. {\bf 81}, 011006 (2012).

\bibitem{Nat_Hashimoto2014} Makoto Hashimoto, Inna M. Vishik, Rui-Hua He, Thomas P. Devereaux and Zhi-Xun Shen, Energy gaps in high-transition-temperature cuprate superconductors, Nature Physics {\bf 10}, 483 (2014).

\bibitem{JPhysSocJap_Nakano1998} Tohru Nakano, Naoki Momono, Migaku Oda, and Masayuki Ido, Correlation between the Doping Dependences of Superconducting Gap Magnitude $2\Delta_0$ and Pseudogap Temperature $T^*$ in High-T$_c$ Cuprates , J. Phys. Soc. Jpn. {\bf 67}, 2622-2625 (1998).

\bibitem{RepProgPhys_Vishik2018} I.M. Vishik, Photoemission perspective on pseudogap, superconducting fluctuations, and charge order in cuprates: a review of recent progress. Reports On Progress in Physics. Physical Society (Great Britain) {\bf 81}, 062501 (2018).

\bibitem{PRB_Zhong2018} Y.-G. Zhong, J.-Y. Guan, X. Shi, J. Zhao, Z.-C. Rao, C.-Y. Tang, H.-J. Liu, Z. Y. Weng, Z. Q. Wang, G. D. Gu, T. Qian, Y.-J. Sun, and H. Ding,
Continuous doping of a cuprate surface: Insights from in situ angle-resolved photoemission
Phys. Rev. B {\bf 98}, 140507(R) (2018).

\bibitem{PRB_Cyr-Choiniere2018} J. Chang, B. J. Ramshaw, D. A. Bonn, W. N. Hardy, R. Liang, J.--Q. Yan, J.-G. Cheng, J.--S. Zhou, J. B. Goodenough, S. Pyon, T. Takayama, H. Takagi, N. Doiron-Leyraud, and Louis Taillefer, Pseudogap temperature $T^*$ of cuprate superconductors from the Nernst effect O. Cyr-Choini\`re, R. Daou, F. Lalibert\'e, C. Collignon, S. Badoux, D. LeBoeuf,  Phys. Rev. B {\bf 97}, 064502 (2018).

\bibitem{PhysicaC_Naqib2003} S. H. Naqib, J. R. Cooper, J. L. Tallon, C. Panagopoulos, Temperature dependence of electrical resistivity of high--$T_c$ cuprates--from pseudogap to overdoped regions, Physica C  {\bf 387},365 (2003).

\bibitem{PNAS_Barisic2013} Neven Barisic, Mun K Chan, Yuan Li, Guichuan Yu, Xudong Zhao, Martin Dressel, Ana Smontara, Martin Greven, Universal sheet resistance and revised phase diagram of the cuprate high-temperature superconductors, Proc Natl Acad Sci {\bf 110}, 12235-40 (2013).

\bibitem{NatCom_Sterpetti2017} E. Sterpetti, J. Biscaras, A. Erb and A. Shukla,
Comprehensive phase diagram of two-dimensional space charge doped Bi$_2$Sr$_2$CaCu$_2$O$_{8+x}$, Nat. Commun. {\bf 8}, 2060 (2017).

\bibitem{PhysicaC_Tallon2001} J. L. Tallon and J.W. Loram, The doping dependence of $T^*$ - what is the real high-$T_c$ phase diagram?, Physica C {\bf 349}, 53 (2001).

\bibitem{ScSciTech_Naqib2008} S. H. Naqib and R. S. Islam, Extraction of the pseudogap energy scale from the static magnetic susceptibility of single and double CuO$_2$ plane high-T$_c$ cuprates,Supercond. Sci. Technol. {\bf 21}, 105017 (2008).

\bibitem{PhysSolStat_Tallon1999} J. L. Tallon, J. W. Loram, G. V. M. Williams, J. R. Cooper, I. R. Fisher, J.D. Johnson, M.P. Staines and C Bernhard,
Critical Doping in Overdoped High-Tc Superconductors: a Quantum Critical Point?, phys. stat. sol. (b) {\bf 215}, 531 (1999).

\bibitem{Solstatcom_Noat2022}  Y. Noat, A. Mauger, M. Nohara, H. Eisaki, W. Sacks, Cuprates phase diagram deduced from magnetic susceptibility: what is the 'true' pseudogap line?, Solid State Communications {\bf 348--349}, 114689 (2022).

\bibitem{PRL_Johnston1989} David C. Johnston, Magnetic Susceptibility Scaling in La$_{2-x}$Sr$_x$CuO$_{4-y}$, Phys. Rev. Lett. {\bf 62}, 957 (1989).

\bibitem{PRB_Torrance1989} J. B. Torrance, A. Bezinge, A. I. Nazzal, T. C. Huang, S. S. P. Parkin, D. T. Keane, S. J. LaPlaca, P. M. Horn, and G. A. Held, Properties that change as superconductivity disappears at high-doping concentrations in La$_{2-x}$Sr$_x$CuO$_4$, Phys. Rev. B {\bf 40}, 8872 (1989).

\bibitem{PhysicaC_Yoshizaki1990} R. Yoshizaki, N. Ishikawa, H. Sawada, E. Kita, A. Tasaki, Magnetic susceptibility of normal state and superconductivity of La$_{2-x}$Sr$_x$CuO$_4$, Physica C {\bf 166}, 417 (1990).

\bibitem{PhysicaC_Oda1991} M. Oda, T. Nakano, Y. Kamada, M. Ido, Electronic states of doped holes and magnetic properties in La$_{2-x}$M$_x$CuO$_4$ (M = Sr, Ba), Physica C {\bf 183}, 234 (1991).

\bibitem{PRB_Nakano1994} T. Nakano, M. Oda, C. Manabe, N. Momono, Y. Miura, and M. Ido, Magnetic properties and electronic conduction of superconducting  La$_{2-x}$Sr$_x$CuO$_4$, Phys. Rev. B {\bf 49}, 16000 (1994).

\bibitem{JphysChemSol_Lines1970} M. E. Lines, The quadratic-layer antiferromagnet, J. Phys. Chem. Solids . {\bf 31}, 101 (1970).

\bibitem{JPhysSocJp_Suda2016} Tomoharu Suda and Takao Morinari, Destruction of Magnetic Long--Range Order by Hole-Induced Skyrmions in Two-Dimensional Heisenberg Model, J. Phys. Soc. Jpn. {\bf 85}, 114702 (2016).

\bibitem{EPL_Sacks2017} W. Sacks, A. Mauger and Y. Noat, Cooper pairs without glue in high-$T_c$ superconductors: A universal phase diagram, Euro. Phys. Lett {\bf 119}, 17001 (2017).

\bibitem{PRB_Birgeneau1988}
R. J. Birgeneau, D. R. Gabbe, H. P. Jenssen, M. A. Kastner, P. J. Picone, T. R. Thurston, G. Shirane, Y. Endoh, M. Sato, K. Yamada, Y. Hidaka, M. Oda, Y. Enomoto, M. Suzuki, and T. Murakami, Antiferromagnetic spin correlations in insulating, metallic, and superconducting La$_{2-x}$Sr$_x$CuO$_4$, Phys. Rev. B {\bf 38}, 6614 (1988).


\bibitem{RevModPhys_Mabousakis1991} Efstratios Manousakis, The spin-1/2 Heisenberg antiferromagnet on a square lattice and its application to the cuprous oxides, Rev. Mod. Phys. {\bf 63}, 1 (1991).

\bibitem{AnRevCondMat_Proust2019} C. Proust and L. Taillefer, The Remarkable Underlying Ground States of Cuprate, Annual Review of Condensed Matter Physics {\bf 10}, 409 (2019).

\bibitem{PRB_Singh1992} Ravi P. Singh, Z. C. Tao, and M. SinghRavi P. Singh, Z. C. Tao, and M. Singh, Role of antiferromagnetic interlayer coupling on magnetic properties of
YBa$_2$Cu$_3$O$_{6+x}$, Phys. Rev. B {\bf 46}, 1244 (1992).

\bibitem{PRB_LeTacon2013} M. Le Tacon, M. Minola, D. C. Peets, M. Moretti Sala, S. Blanco-Canosa, V. Hinkov, R. Liang, D. A. Bonn, W. N. Hardy, C. T. Lin, T. Schmitt, L. Braicovich, G. Ghiringhelli, and B. Keimer, Dispersive spin excitations in highly overdoped cuprates revealed by resonant inelastic x-ray scattering, Phys. Rev. B {\bf 88}, 020501(R) (2013).

\bibitem{PRB_Peng2018} Y. Y. Peng, E. W. Huang, R. Fumagalli, M. Minola, Y. Wang, X. Sun, Y. Ding, K. Kummer, X. J. Zhou, N. B. Brookes, B. Moritz, L. Braicovich, T. P. Devereaux, and G. Ghiringhelli, Dispersion, damping, and intensity of spin excitations in the monolayer (Bi,Pb)$_2$(Sr,La)$_2$CuO$_{6+\delta}$ cuprate superconductor family, Phys. Rev. B {\bf 98}, 144507 (2018).

\bibitem{Nature_Dean2013} M. P. M. Dean, G. Dellea, R. S. Springell, F. Yakhou-Harris, K. Kummer, N. B. Brookes, X. Liu, Y-J. Sun, J. Strle, T. Schmitt, L. Braicovich, G. Ghiringhelli, I. Bov\v{z}ovi\'{c} and J. P. Hill, Persistence of magnetic excitations in  La$_{2-x}$Sr$_x$CuO$_4$ from the undoped insulator to the heavily overdoped non-superconducting metal, Nature Materials {\bf 12}, 1019 (2013).

\bibitem{SciTech_Sacks2015}  W. Sacks, A. Mauger, Y. Noat, Pair\,--\,pair interactions as a mechanism for
high-T$_c$ superconductivity, Superconduct. Sci. Technol., {\bf 28}
105014 (2015).

\bibitem{EPL_StJean2001} M. Saint Jean, C. Even and C. Guthmann, Macroscopic 2D Wigner islands, Eur. Phys. Lett., {\bf 55}, 45 (2001).

\bibitem{Jphys_Sacks2018}W. Sacks, A. Mauger and Y. Noat, Origin of the Fermi arcs in cuprates: a dual role of quasiparticle and pair excitations, Journal of Physics: Condensed Matter, {\bf 30},  475703 (2018).

\bibitem{SolStatCom_Noat2021} Y. Noat, A. Mauger, M. Nohara, H. Eisaki, W. Sacks
, How `pairons' are revealed in the electronic specific heat of cuprates, Solid State Communications {\bf 323}, 114109 (2021).

\bibitem{PNAS_Vishik2012} I. M. Vishik, M. Hashimoto, R.-H. He, W.-S. Lee, F. Schmitt, D. Lu, R. G. Moore, C. Zhang, W. Meevasana, T. Sasagawa, S. Uchida, Kazuhiro Fujita, S. Ishida, M. Ishikado, Y. Yoshida, H. Eisaki, Z. Hussain, T. P. Devereaux, and Z.-X. Shen, Phase competition in trisected superconducting dome, PNAS {\bf 109}, 18332 (2012).

\bibitem{PRB_Liang2006} R. Liang, D. A. Bonn, and W. N. Hardy, Evaluation of
CuO$_2$ plane hole doping in YBa$_2$Cu$_3$O$_{6+x}$ single crystalsPhys. Rev. B {\bf 73}, 180505 (2006).

\bibitem{EPJB_StJean2004} M. Saint Jean, C. Guthmann and G. Coupier, Relaxation and ordering processes in``macroscopic Wigner crystals'', Eur. Phys. J. B {\bf 39}, 61-68 (2004).

\bibitem{Form_Tanemura2003} M. Tanemura, Statistical distributions of Poisson Voronoi cells in two and three dimensions, Forma {\bf 18}, 221 (2003).

\bibitem{PhysicaA_Szabo2007}J\'arai-Szab\'o Ferenc, Zolt\'an N\'eda, On the size--distribution of Poisson Voronoi cells,  Physica A, {\bf 385}, pp 518-526 (2007).

\bibitem{PRL_Pan2000} S. H. Pan, E. W. Hudson, A. K. Gupta, K.-W. Ng, H. Eisaki, S. Uchida, and J. C. Davis, STM Studies of the Electronic Structure of Vortex Cores in Bi$_2$Sr$_2$CaCu$_2$O$_{8+\delta}$, Phys. Rev. Lett. {\bf 85}, 1536 (2000).

\bibitem{PRB_Sacks2006} W. Sacks, T. Cren, D. Roditchev, and B. Dou\c{c}ot, Quasiparticle spectrum of the cuprate Bi$_2$Sr$_2$CaCu$_2$O$_{8+\delta}$: Possible connection to the phase diagram, Phys. Rev. B  {\bf 74}, 174517(2006).

\bibitem{Jphys_Sacks2017} William Sacks, Alain Mauger and Yves Noat, Universal spectral signatures in pnictides and cuprates: the role of quasiparticle-pair coupling, J. Phys.: Condens. Matter  {\bf 29}, 445601 (2017).

\bibitem{EPL_Cren2000} T. Cren, D. Roditchev, W. Sacks and J. Klein, Constraints on the quasiparticle density of states in high-Tc superconductors, , Europhys. Lett. {\bf 52}, 203 (2000).

\bibitem{PRL_Dynes1978}  R. C. Dynes, V. Narayanamurti, and J. P. Garno, Direct Measurement of Quasiparticle-Lifetime Broadening in a Strong-Coupled Superconductor,, Phys. Rev. Lett. {\bf 41}, 1509 (1978).

\bibitem{PRL_Cren2000} T. Cren, D. Roditchev, W. Sacks, J. Klein, J.-B. Moussy, C. Deville-Cavellin, and M. Lagu\'es, Influence of Disorder on the Local Density of States in High-$T_c$ Superconducting Thin Films, Phys. Rev. Lett. {\bf 84}, 147 (2000).

\bibitem{PRB_Howald2001} C. Howald, P. Fournier, and A. Kapitulnik, Inherent inhomogeneities in tunneling spectra of Bi$_2$Sr$_2$CaCu$_2$O$_{8-x}$ crystals in the superconducting state, Phys. Rev. B {\bf 64}, 100504(R) (2001).

\bibitem{Corresp} $^*$ Corresponding author.\\
E-mail address: yves.noat@insp.jussieu.fr (Y. Noat).

\end{thebibliography}
\end{document}